\documentclass[10pt]{article}
\usepackage[margin=1.0in]{geometry}
\usepackage{setspace}
\usepackage{titling}

\usepackage[utf8]{inputenc}

\usepackage{amsmath}
\usepackage{amssymb}
\usepackage{amsthm}
\usepackage{bbm}
\usepackage{mathrsfs}

\newtheorem{assumption}{Assumption}
\newtheorem{definition}{Definition}
\newtheorem{lemma}{Lemma}
\newtheorem{theorem}{Theorem}
\newtheoremstyle{named}{}{}{\itshape}{}{}{\textbf{.}}{.5em}{\thmnote{#3}}
\theoremstyle{named}
\newtheorem*{namedtheorem}{Theorem}

\usepackage{subcaption}
\usepackage[nolists]{endfloat}
\usepackage{tikz}
\usepackage{pgfplots}
\usepackage{subfiles}
\graphicspath{{Figures/}}

\usepackage[capitalise,noabbrev]{cleveref}
\crefname{assumption}{Assumption}{Assumptions}
\crefname{definition}{Definition}{Definitions}
\crefname{lemma}{Lemma}{Lemmata}
\crefname{theorem}{Theorem}{Theorems}
\usepackage[sort]{natbib}
\numberwithin{equation}{section}
\allowdisplaybreaks

\def\ci{\perp\!\!\!\perp}
\DeclareMathOperator*{\argmin}{argmin}

\title{Stochastic treatment choice with empirical welfare updating\thanks{%
The authors thank participants at the Bristol Econometrics Study Group and Midwest Econometrics Group conferences for beneficial comments. 
The authors also gratefully acknowledge financial support from the ERC (grant number 715940) and the ESRC Centre for Microdata Methods and Practice (CeMMAP; grant number RES-589-28-0001).}}
\date{\today}
\author{
        Toru Kitagawa\thanks{Department of Economics, Brown University. Email: toru\_kitagawa@brown.edu}
            \and
        Hugo Lopez\thanks{Department of Economics, University of Chicago.}
            \and
        Jeff Rowley\thanks{Department of Economics, University College London.}
}

\begin{document}
\maketitle

\begin{abstract}
    This paper proposes a novel method to estimate individualised treatment assignment rules. 
    The method is designed to find rules that are stochastic, reflecting uncertainty in estimation of an assignment rule and about its welfare performance. 
    Our approach is to form a prior distribution over assignment rules, not over data generating processes, and to update this prior based upon an empirical welfare criterion, not likelihood.
    The social planner then assigns treatment by drawing a policy from the resulting posterior. 
    We show analytically a welfare-optimal way of updating the prior using empirical welfare; 
    this posterior is not feasible to compute, so we propose a variational Bayes approximation for the optimal posterior. 
    We characterise the welfare regret convergence of the assignment rule based upon this variational Bayes approximation, showing that it converges to zero at a rate of $\ln\left(n\right)/\sqrt{n}$.
    We apply our methods to experimental data from the Job Training Partnership Act Study to illustrate the implementation of our methods. 
    \vspace{\baselineskip}
    
\end{abstract}


\section{Introduction}
\label{SEC:INTRODUCTION}

The principal goal of programme evaluation is to inform the social planner as to which individuals within a target population should receive which treatment. 
When treatment effects are heterogeneous in individuals' observable characteristics, the social planner can improve social welfare by implementing an individualised treatment assignment rule based upon these characteristics. 
The literature on statistical treatment choice initiated by \citet{Manski2004} studies how to estimate assignment rules based upon a finite sample and how to assess their welfare performance.
Given an experimental or observational sample, existing approaches -- including those proposed in \citet{athey2021policy}, \citet{Hirano2009}, \citet{Kitagawa2018a}, and \citet{Manski2004} -- yield deterministic assignment rules, which are functions mapping the individual's observable characteristics to a recommended treatment. 
That is, individuals that share the same observable characteristics are all assigned the same treatment. 
Such assignment rules are sharp and address the question of \textit{who should be treated?}
We adopt a broader view of the treatment choice problem by considering \textit{stochastic assignment rules} that map individual observable characteristics to a probability distribution over the different treatment arms, instead addressing the question of \textit{with what probability should an individual be treated?}

In static treatment choice problems with outcome distributions that exhibit the monotone likelihood ratio property, deterministic assignment rules form a class of admissible policies \citep{karlin1956theory,tetenov2012statistical}, such that restricting attention to this class is without loss (of welfare).
Once we allow the class of outcome distributions to be unconstrained though, there is little theoretical justification for focusing on deterministic rules. 
In comparison to stochastic assignment rules, deterministic assignment rules have the following three potentially undesirable features. 
First, deterministic assignment rules cannot incorporate  confidence or uncertainty about which treatment is best for each individual, with individuals typically assigned treatment if conventional point estimates suggest that treatment has a positive effect on average.
The strength of evidence in support of this conclusion, usually presented in the form of confidence intervals or p-values, is not generally acted upon;
what matters is whether empirical evidence supports a positive point estimate, and not
whether it is sufficient or insufficient to reject a non-positive effect.
Such a sharp dichotomy of assignment is naturally overconfident in its prescription, and there is no theoretical basis for the incorporation of confidence intervals or p-values into frequentist-based decision-making.
Stochastic assignment rules can represent such uncertainty, which arises due to the (finite sample) nature of experimental data or model misspecification, through their probability weighting of treatments. 
Second, stochastic assignment rules facilitate future evaluation, since implementing a stochastic assignment rule can generate a new experimental sample in which treatment is randomised conditional on individual observable characteristics. 
Third, unlike deterministic assignment rules for which the probability that a treatment is assigned changes discontinuously at some threshold, stochastic assignment rules feature assignment probabilities that smoothly change with respect to individual characteristics. 
Such a feature is desirable if a fairness criterion requiring that individuals with similar characteristics have similar probabilities of treatment \citep{dwork2012fairness} is enforced.

This paper proposes novel and general methods for obtaining stochastic individualised assignment rules based on randomised control trial data.
Assuming that the social planner assigns individuals to a binary treatment, with her goal being to maximise additive (utilitarian) social welfare as in \citet{Manski2004}, we exploit an empirical analogue of the social welfare criterion to generate individualised assignment probabilities.
Specifically, we start with a prior distribution over a collection of deterministic assignment rules $\mathscr{G}$, each of which partitions the space of individual observable characteristics $\mathscr{X}$ into a group of characteristics $G$ and its complement $G^{c}$, thereby generating a deterministic assignment rule $g\left(x\right)=1\left( x \in G \right)$.
An individual is assigned treatment if their characteristics are such that $x\in G$, and is not assigned treatment if $x\in G^{c}$.
We then update the prior distribution based upon an empirical analogue of the social welfare criterion to obtain a posterior distribution over $\mathscr{G}$. 
To generate a stochastic assignment, we draw a $g\in\mathscr{G}$ according to the posterior distribution over this collection, and implement the policy prescribed by $g$.
In this way, the stochastic assignment ruleapplied to an individual with characteristics $x$ treats her with the probability equal to the posterior probability that stochastic $G$ contains $x$. 

One of the main contributions of this paper is that we derive an optimal updating procedure for obtaining the posterior distribution over $\mathscr{G}$. 
This procedure minimises an upper bound on welfare regret and yields an exponential tilting of the prior over $\mathscr{G}$, where the tilting depends upon the empirical welfare criterion. 
This novel updating formula resembles the quasi-posterior distribution that appears in the Laplace-type estimation studied by \citet{Chernozhukov03}, but differs in that the constant factor in the exponential tilting term is determined endogenously by the Lagrange multiplier of the optimisation. 

Despite our analytical characterisation of the optimal posterior distribution, computation of this distribution or sampling of $g$ from it is not straightforward.
We therefore consider a variational approximation of the optimal posterior distribution by a parametric distribution. 
In particular, as a specification of $\mathscr{G}$, we consider the class of Linear Eligibility Score rules that assign treatment if $x^{\top}\gamma$ (the linear score) exceeds some threshold $c$ (eligibility). 
Building upon the LES class, we exploit the invariance of the welfare criterion to the ratio of $\gamma$ to $c$ (scale invariance) and approximate the optimal posterior distribution using a multivariate von Mises-Fisher distribution. 

For the practice of reporting and communicating individualised allocations of treatment, our approach of obtaining a posterior distribution over policies is useful for generating some quantities that existing methods yielding deterministic assignment rules cannot produce. 
First, the posterior probability that $g(x)=1$ offers a personalised probabilistic assessment that individuals with characteristics $x$ favour treatment over no treatment. 
Reporting such a probability offers a novel alternative to the common practice of using the p-values of hypothesis testing to express confidence in a positive treatment effect, something which does not easily translate to a recommendation about what the social planner should do.
Second, viewing the posterior over $g$ as an inferential tool for the welfare-optimality of (deterministic) assignment policies within $\mathscr{G}$, we can obtain a credible region for the optimal policy by, for instance, selecting its highest posterior density region. 
This approach to obtaining confidence sets for the optimal treatment assignment policy is an alternative to the frequentist approach that is studied in \citet{rai2019}.
Third, analogous to the practice of using a Bayesian posterior with a noninformative prior as a visual summary of the likelihood function, we can use our variationally approximated posterior over $g$ as a visual summary of the exponentiated empirical welfare criterion function.

To demonstrate how to implement our approach and what it delivers in practice, we apply our methods to the JTPA Study sample that is studied by \citet{bloom1997benefits}.
Given observations of prior earnings and years of education, we ask \textit{with what probability should an individual be treated?}
Restricting attention to linear assignment rules and a variational approximation of the optimal posterior distribution by a multivariate von Mises-Fisher distribution, we estimate a stochastic assignment rule that is more likely to allocate JTPA assistance to individuals with high prior earnings and fewer years of education.
\citet{Kitagawa2018a}, which similarly considers the JTPA Study sample, and estimates a deterministic assignment rule, serves as a useful benchmark for comparison. 
Aside from the obvious difference that we estimate a stochastic rule (i.e., every individual has a non-trivial probability to be allocated JTPA assistance under our rule), our estimated rule allocates JTPA assistance to a smaller fraction of the population than the deterministic rule of \citet{Kitagawa2018a}, which targets individuals with low prior earnings and few years of education for treatment. 
This difference reflects the shape of the empirical welfare criterion, which can be captured by our approach but is missed by deterministic policies that are obtained as the mode of the empirical welfare criterion.


\subsection{Literature review}
\label{SEC:LIT}

This paper contributes to the growing literature on statistical treatment choice initiated by \citet{Manski2004}.
Exact minimax regret assignment rules are studied in \citet{ishihara2021}, \citet{schlag2006eleven}, \citet{Stoye2009}, \citet{Stoye2012}, \citet{tetenov2012statistical}, and \citet{yata2021}.
\citet{Hirano2009} analyses asymptotically-optimal assignment rules in limit experiments, and \citet{bhattacharya2012} considers capacity constrained policies.  
\citet{Kitagawa2018a} proposes \textit{Empirical Welfare Maximisation} (EWM) methods for individualised assignment, which maximise a sample analogue of the social welfare function over a constrained class of policies.
Similar approaches have been studied in the literature on machine learning and personalised medicine, as in \citet{Beygelzimer2009}, \citet{Swaminathan2015}, \citet{Zadrozny2003}, \citet{Zhang2012}, and \citet{Zhao2012}. 
Recent advances in learning individualised assignment policies include \citet{adjaho2022externally}, \citet{athey2021policy}, \citet{dadamo21}, \citet{han2022}, \citet{kido22}, \citet{Kitagawa2017}, \citet{KST21}, \citet{liu22}, \citet{Mbakop2018}, \citet{nie2019learning}, \citet{Sakaguchi2019},  \citet{sasaki20}, \citet{sun21}, and \citet{viviano21}, to list but a few works. 
The assignment rules estimated in these works are all deterministic. 

There are some earlier works that investigate the decision-theoretic justification for stochastic (fractional) assignment and the welfare performance of these rules, with \citet{manski20092009} providing a detailed review of settings where stochastic rules are preferable. 
When the welfare criterion is only partially identified, minimax regret-optimal rules are stochastic given knowledge of the identified set \citep{Manski2000,manski2005social,manski2007identification, Manski2007}, which remains true even after taking into account uncertainty of estimates of the bounds \citep{Stoye2012, yata2021, manski2022identification}. 
As shown by \citet{manski2007admissible} and \cite {manski20092009}, stochastic assignment rules can also be justified by a nonlinear welfare criterion in a point-identified setting. 
\citet{kitagawa2022} shows that, for a wide class of nonlinear welfare regret criteria, admissible assignment rules are stochastic (fractional). 
In particular, \citet{kitagawa2022} shows that the minimax squared-regret rule is stochastic, with the probability of assignment equal to the posterior probability of a positive treatment effect under the least-favorable null. 
\citet{kitagawa2022} proposes using this probability as a measure of the strength of evidence for a positive treatment effect, replacing the commonly used p-value of a hypothesis test. 
In contrast, this paper obtains the probability of assignment from a posterior probability distribution over assignment rules, rather than over the treatment effect parameters, with a quasi-likelihood built upon the empirical welfare criterion.
\citet{kock2022treatment} obtains a stochastic assignment rule in a setting where the oracle optimal rule is fractional due to nonlinearity in the social planner's chosen welfare criterion.
In contrast to the static treatment assignment problem, dynamic treatment assignment problems analysed in the multi-arm bandit literature often consider stochastic assignments that balance the exploitation versus exploration trade-off, such as the posterior probability matching algorithm of \citet{thompson1933likelihood} does. Thompson sampling algorithms build upon the standard Bayesian posterior distribution for treatment effects such that the allocation algorithm crucially relies on a parametric specification of the data generating process, which our approach does not require.

\citet{Chamberlain2011} and \citet{Dehejia2005} approach the treatment choice problem from a Bayesian perspective.
In their framework, the potential outcome distributions are parametric, and it is over the parameters of these distributions that a prior is imposed. 
For the standard mean welfare criterion, the Bayes optimal allocation rule is deterministic.
Our approach differs from these works in that we do not assume a prior distribution over the data generating process. 
We instead impose few restrictions on the data generating process, and form prior and posterior distributions over the parameters that index assignment rules. 
Our approach can be advantageous when compared to \citet{Chamberlain2011} if the social planner is concerned about potential misspecification of the likelihood.
If the likelihood is misspecified, the resulting Bayes-optimal assignment rule can be suboptimal even for large samples. 
In contrast, our approach is guaranteed to yield a distribution over policies that is guaranteed to concentrate on welfare-optimal policies without requiring a specification for the data generating process.   

Our approach is also related to that of \citet{Bissiri2016} and \citet{csaba2020learning}, where loss function-driven (quasi-Bayes) updating rules are proposed. 
Rather than follow their approach by adopting exponentiated loss as a quasi-likelihood and solving the quasi-Bayesian decision problem, we obtain an optimal learning rule by minimising a high probability upper bound on welfare regret. 
This way of establishing optimality is similar to the structural risk minimisation approach of \citet{Vapnik1998} and the \emph{Probably Approximately Correct} (henceforth, PAC) analysis proposed by \citet{Valiant1984}, which was extended to the study of randomised predictors in \citet{McAllester1999}, and \citet{Shawe-Taylor1997}, constituting the development of PAC-Bayes theory.
For classification and regression problems, various PAC-Bayes bounds on prediction generalisation errors are obtained in \citet{Begin2014,Begin2016}, \citet{catoni2007pac}, \citet{Derbeko2004}, \citet{Germain2009}, \citet{McAllester2003}, \citet{Pentina2015}, and \citet{Seeger2002}, and can accommodate quasi-Bayesian procedures similar to ours.
See \citet{Guedj2019} for a recent review of this literature. 
To our knowledge, the PAC-Bayes bounds that we derive for treatment choice are new to the literature and offer a contribution of independent interest. We also note that \citet{pellatt2022pac} makes use of PAC-Bayes theory to analyse the treatment allocation problem with stochastic assignment rules under a budget (or resource) constraint. 

Although the treatment choice problem is distinct from prediction problems -- as is discussed in \citet{Kitagawa2018a} -- the EWM approach for treatment choice is closely related to the cost-sensitive binary \emph{classification} problem, as first pointed out by \citet{Zadrozny2003}. 
The PAC-Bayes classification analysis with variational posterior approximation that is proposed by \citet{Alquier2016} is, therefore, closely related to our analysis. 
There are, however, important differences with \citet{Alquier2016}.
First, we make use of the approach proposed by \citet{Begin2016}, which allows for the construction of a variety of different bounds via a general convex function. 
This introduces the complication that classification is not standard (i.e., cost is homogeneous) and is instead cost-sensitive.
Introducing heterogeneity in the cost of misclassification leads to a non-trivial challenge in deriving the PAC bounds. 
To address these complications we leverage results in \citet{Maurer2004} for continuous loss functions over the unit interval. 
Second, \citet{Alquier2016} considers approximating the optimal posterior distribution by a Gaussian distribution. 
We exploit the scale invariance property of the welfare criterion and approximate the optimal posterior distribution by a multivariate von Mises-Fisher distribution over the hypersphere.


\section{Model and Setup} 
\label{SEC:MODEL}


\subsection{Notation and setting}
\label{SEC:NOTATION}

We let $D \in \{1,0\}$ be a binary treatment; 
we let $Y_{1}\in\mathbb{R}$ and $Y_{0}\in\mathbb{R}$ be the potential outcomes associated with the two treatment states; 
and we let $X\in\mathscr{X}$ be a vector of observed characteristics, which we refer to as covariates, where $\mathscr{X}\subset\mathbb{R}^{m-1}$.
We suppose that experimental data comprising $n$ independent and identically distributed observations concatenated as $\left(Y_{i},D_{i},X_{i}\right)$ are available, where $Y_{i} = D_{i}\cdot Y_{1i} + \left(1-D_i\right)\cdot Y_{0i}$ is the post-treatment observed outcome of individual $i$.
We denote the joint distribution of the experimental sample, which we reiterate is a probability distribution over $\left(Y_{i},D_{i},X_{i}\right)$, by $P^{n}$. 
The joint distribution of $\left(Y_{1},Y_{0},D,X\right)$ that induces the independent and identical distribution of the observations $\left(Y_{i},D_{i},X_{i}\right)$, $i=1, \dots, n$ is referred to as the data generating process and denoted by $P$.
We assume that the covariates consist only of those characteristics that the planner can use to discriminate between individuals in the target population, with budgetary, ethical or legal considerations precluding the use of other characteristics.

Throughout our analysis, we maintain several assumptions.
We follow \citet{Manski2004} and the subsequent literature in supposing that the social welfare criterion is that of a utilitarian social planner who aims to maximise the average level of individual outcomes. 
We note that other criteria could also be implemented, such as inequality-averse social welfare and Gini social welfare.\footnote{%
For example, \citet{Kasy2016} and \citet{Kitagawa2017} study a setting where the social welfare function is a weighted average of the outcomes with rank-dependent weights, which includes the Gini social welfare function as a special case.} 
\begin{assumption}[External validity]
\label{ASS:EV}
The population to which policy is to be applied -- the target population -- has the same distribution over $\left(Y_{1},Y_{0},X\right)$ as the marginal distribution of $\left(Y_{1},Y_{0},X\right)$ that is obtained from the data generating process.
\end{assumption}
\begin{assumption}[Unconfoundedness]
\label{ASS:U}
The data generating process satisfies $\left(Y_{1},Y_{0}\right)\ci D|X$.
\end{assumption}
\cref{ASS:EV,ASS:U} are satisfied, for instance, if the experimental data is extracted directly from the target population and the treatment is, conditional on the covariates, randomly assigned,\footnote{%
\citet{Kitagawa2018a} considers a setting where the marginal distribution of $X$ differs between the population of interest and the data generating process. 
\citet{adjaho2022externally} and \citet{kido22} study settings that differ also in terms of the distribution of potential outcomes.}
independently of the potential outcomes \citep{Rosenbaum1983}.
\begin{assumption}[Bounded outcomes]
\label{ASS:BO}
There exists a constant $0<M<\infty$ such that the support of Y is contained in $[0,M]$.
\end{assumption}
\begin{assumption}[Strict overlap]
\label{ASS:SO}
There exists a $\psi\in\left(0,1/2\right)$ such that the propensity score satisfies, for all $x\in\mathscr{X}$, $e\left(x\right)\in[\psi,1-\psi]$, where $e\left(x\right)=\mathrm{E}_{P}\left(D|X\right)$.
\end{assumption}
We normalise the outcome variable to the interval $[0,M]$. 
As is discussed in \citet{Kitagawa2018a} and \citet{Swaminathan2015}, policies that maximise an empirical welfare criterion are not invariant to positive affine transformations of outcomes, which is the case for the empirical welfare criterion that we consider in this paper. 
Given \cref{ASS:BO,ASS:SO}, we can define
\begin{equation}
\label{EQ:WEIGHTSFORIMPLEMENTATION}
    H
    \equiv 
    \frac{\psi\cdot Y/M}{e\left(X\right)\cdot D + \left(1-e\left(X\right)\right)\cdot\left(1-D\right)},
\end{equation}
which is confined to the unit interval, and which we interpret as weights and that are motivated by an unbiased estimator of the (scaled) expected potential outcomes.
We solve the planner's problem using these transformed outcomes. This transformation of outcomes does not affect the welfare ranking over assignment policies, both in the population and according to in-sample welfare criteria, yet facilitates the proof of our PAC-Bayes bounds shown in Theorem \ref{THM:PAC} below.

Let $G\subset\mathscr{X}$ specify a set of individuals -- identified by their individual characteristics -- to be treated deterministically.  
We refer to $G$ as a policy.
Adopting an additive utilitarian perspective, the average level of social welfare attained by $G$ is proportional to
\begin{equation}
\label{EQ:ORIGINAL}
    W\left(G\right) 
    \equiv
    \mathrm{E}_{P}\left(Y_{1} \cdot 1\left(X \in G\right)+Y_{0} \cdot 1\left(X \notin G\right)\right).
\end{equation}
Given \cref{ASS:EV,ASS:U,ASS:SO} and that $Y=Y_{0}+D\cdot\left(Y_{1}-Y_{0}\right)$, we can re-write \cref{EQ:ORIGINAL} as
\begin{align}
    W\left(G\right)
    &=
    \mathrm{E}_{P}\left(\frac{Y\cdot D}{e\left(X\right)} \cdot 1\left(X \in G\right)+\frac{Y\cdot \left(1-D\right)}{1-e\left(X\right)} \cdot 1\left(X \notin G\right)\right),\\
    &=
    \mathrm{E}_{P}\left(\frac{Y\cdot\left(1-D\right)}{1-e\left(X\right)}\right)+\mathrm{E}_{P}\left(\left( \frac{Y\cdot D}{e\left(X\right)}-\frac{Y\cdot\left(1-D\right)}{1-e\left(X\right)}\right)\cdot 1\left(X\in G\right)\right). 
\end{align}
Accordingly, the sample analogue of \cref{EQ:ORIGINAL} can be written as
\begin{align}
    W_{n}\left(G\right)
    &\equiv
    \frac{1}{n}\sum^{n}_{i=1}\left(\frac{y_{i}\cdot d_{i}}{e\left(x_{i}\right)} \cdot 1\left(x_{i} \in G\right)+\frac{y_{i}\cdot\left(1-d_{i}\right)}{1-e\left(x_{i}\right)} \cdot 1\left(x_{i} \notin G\right)\right),\\
    &=
    \frac{1}{n}\sum^{n}_{i=1}\frac{y_{i}\cdot\left(1-d_{i}\right)}{1-e\left(x_{i}\right)}+\frac{1}{n}\sum^{n}_{i=1}\left(\frac{y_{i}\cdot d_{i}}{e\left(x_{i}\right)}-\frac{y_{i}\cdot\left(1-d_{i}\right)}{1-e\left(x_{i}\right)}\right)\cdot 1\left(x_{i}\in G\right),
    \label{EQ:TOREWRITE}
\end{align}
where $W_{n}\left(G\right)$ is an unbiased estimator for the true level of welfare that arises from the implementation of a particular $G$. 
Given the additive social welfare criterion, the maximal welfare level can be attained by a deterministic policy. 
Hence, as far as the population welfare maximisation problem is concerned, the social planner wants to select the $G$ that maximises $W\left(G\right)$.

Each $G$ can be associated with a binary function $g$ that indicates membership in $G$. 
We refer to $g$ as a deterministic asssignment rule, or simply as an assignment rule, with $\mathscr{G}$ constituting the class of assignment rules.
With this notation to hand, we can write \cref{EQ:TOREWRITE}, with some abuse of notation, as
\begin{align}
W_n\left(g\right)
&=
\frac{1}{n}\sum^{n}_{i=1}\frac{y_{i}\cdot\left(1-d_{i}\right)}{1-e\left(x_{i}\right)}+\frac{1}{n}\sum^{n}_{i=1}\left(\frac{y_{i}\cdot d_{i}}{e\left(x_{i}\right)}-\frac{y_{i}\cdot\left(1-d_{i}\right)}{1-e\left(x_{i}\right)}\right)\cdot g\left(x_{i}\right),\\
&=
\frac{1}{n}\sum^{n}_{i=1}h_{i}\cdot\frac{M}{\psi}\cdot\left(e\left(x_{i}\right)\cdot d_{i}+\left(1-e\left(x_{i}\right)\right)\cdot\left(1-d_{i}\right)\right)\cdot\left[\frac{d_{i}}{e\left(x_{i}\right)}\cdot g\left(x_{i}\right)+\frac{1-d_{i}}{1-e\left(x_{i}\right)}\cdot\left(1-g\left(x_{i}\right)\right)\right],\\
&=
\frac{1}{n}\sum^{n}_{i=1}h_{i}\cdot\frac{M}{\psi}\cdot\left[d_{i}\cdot g\left(x_{i}\right)+\left(1-d_{i}\right)\cdot\left(1-g\left(x_{i}\right)\right)\right],\\
&=
\frac{1}{n}\sum^{n}_{i=1}h_{i}\cdot\frac{M}{\psi}\cdot 1\left(g\left(x_{i}\right)=d_{i}\right),\\
&=
\frac{1}{n}\sum^{n}_{i=1}h_{i}\cdot\frac{M}{\psi}-\frac{1}{n}\sum^{n}_{i=1}h_{i}\cdot\frac{M}{\psi}\cdot 1\left(g\left(x_{i}\right)\neq d_{i}\right),
\label{EQ:WEIGHT}
\end{align}
where $h_i$ is the realisation of $H$, as defined in \cref{EQ:WEIGHTSFORIMPLEMENTATION}, for observation $\left(y_{i},d_{i},x_{i}\right)$.
We observe from \cref{EQ:WEIGHT} that $W_{n}\left(g\right)$ depends upon $g$ only through its second term, such that
\begin{equation}
\label{EQ:OBJECTIVE}
    \underset{g}{\operatorname{argmax}}\,W_n\left(g\right) = \underset{g}{\operatorname{argmin}}\,\frac{1}{n}\sum_{i=1}^n h_{i}\cdot 1\left( g\left(x_{i}\right) \neq d_{i} \right).
\end{equation}
Accordingly, we define
\begin{equation}
    R\left(g\right) 
    \equiv
    \mathrm{E}_{P} \left(H\cdot 1\left(g\left(X\right) \neq D\right)\right),
    \label{EQ:LESRISK}
\end{equation}
which we term the \textit{welfare risk} of $g$, and its empirical analogue
\begin{equation}
    R_{S}\left(g\right) 
    \equiv
    \frac{1}{n} \sum_{i=1}^{n}h_{i} \cdot 1\left(g\left(x_{i}\right) \neq d_{i}\right),
    \label{EQ:EMPIRICALLESRISK}
\end{equation}
which we term the \textit{empirical welfare risk} of $g$.
In view of \cref{EQ:OBJECTIVE}, the social planner's objective is to minimise \cref{EQ:LESRISK} via \cref{EQ:EMPIRICALLESRISK} in $g$, following the empirical risk minimisation principle of \cite{Vapnik1998}.

One special set of policies that we draw particular attention to is the \textit{Linear Eligibility Score} (LES) class that is defined in \citet{Kitagawa2018a}, and that we denote by $\mathscr{F}$.
Assignment rules in this class are indexed by a finite-dimensional parameter vector $\gamma$ and a threshold $c$, and are associated with a binary function $f_{\beta}$ that satisfies, for all $x\in\mathscr{X}$,
\begin{equation}
\label{EQ:UNIQUENESSOFINDEX}
    f_{\beta}\left(x\right)
    \equiv
    1\left(x^{\top}\gamma\geq c\right),
\end{equation}
where we take $\beta$ to include both $\gamma$ and $c$ (i.e., $\beta$ is an $m$-dimensional vector).
Each LES rule induces a partitioning of the covariate space into two half-spaces, such that individuals in the upper contour set receive treatment and individuals in the lower contour set do not.
By restricting $\beta$ to the unit hypersphere (i.e., the Euclidean length of $\beta$ is one), we guarantee that each policy is associated with a unique $\beta$.
In what follows, we exploit the interchangeability of $\beta$ and the LES rule that it indexes, adopting $\beta$ as the argument of the loss functions that we consider. 
For instance, and again with some abuse of notation, whenever we focus on the LES class of decision rules we write
\begin{equation}
    R\left(\beta\right)
    =
    \mathrm{E}_{P}\left(H\cdot 1\left(f_{\beta}\left(X\right)\neq D\right)\right),
\end{equation}
and
\begin{equation}
    R_{S}\left(\beta\right) 
    =
    \frac{1}{n} \sum_{i=1}^{n}h_{i} \cdot 1\left\{f_{\beta}\left(x_{i}\right) \neq d_{i}\right\},
\end{equation}
respectively, in place of \cref{EQ:LESRISK,EQ:EMPIRICALLESRISK}.


\subsection{Posterior over policies as a stochastic assignment rule}
\label{SEC:POSTERIOR}

We now adapt \cref{EQ:LESRISK,EQ:EMPIRICALLESRISK}  to handle stochastic assignment rules.
We let $\Pi$ denote a probability distribution over $\mathscr{G}$ that we interpret as a posterior distribution, assuming that $\mathscr{G}$ can be embedded in a measurable space.\footnote{%
For $\mathscr{G}$ to be embedded in a measurable space, $\mathscr{G}$ cannot be too rich. We defer to \citet{molchanov2005} and \citet{gunsilius2019path} for further discussion of this point.}
We let $\mathscr{M}$ denote the collection of all posterior distributions. 
\begin{definition}[Posterior assignment rule]
Let $\Pi$ be a probability distribution over $\mathscr{G}$ that is constructed upon observing the sample. 
The posterior assignment rule under $\Pi$ is a stochastic assignment rule that assigns individuals with $x\in\mathscr{X}$ to treatment with probability $Q^{\Pi}\left(x\right)\equiv\int_{\mathscr{G}}g\left(x\right)\cdot\mathrm{d}\Pi$.
\end{definition}
To implement posterior assignment rules in practice, we randomly draw a $g$ from $\mathscr{G}$ according to $\Pi$ for each individual in the target population.
In this way, similar individuals, who can have similar assignment probabilities, can be assigned to different treatment arms.
Moreover, this approach does not require computation of the probability of treatment.
\begin{definition}[Expected welfare risk under $\Pi$]
We define the expected welfare risk under $\Pi$ as
\begin{equation}
\begin{aligned}
R^{\Pi} 
&\equiv 
\int_\mathscr{G}R\left(g\right)\cdot\mathrm{d}\Pi\left(g\right),\\
&= 
\mathrm{E}_{P} \left(H \cdot\left(D\cdot\left(1-Q^{\Pi}\left(X\right)\right)+ \left(1-D\right)\cdot Q^{\Pi}\left(X\right)\right) \right),
\end{aligned}
\end{equation} 
with its empirical analogue taking the form 
\begin{equation}
\begin{aligned}
    R_{S}^\Pi 
    &\equiv 
    \int_\mathscr{G}R_{S}\left(g\right)\cdot\mathrm{d}\Pi\left(g\right),\\ 
    &= 
    \frac{1}{n}\sum_{i=1}^{n} h_{i} \cdot\left(d_{i}\cdot\left(1-Q^{\Pi}\left(x_{i}\right)\right)+ \left(1-d_{i}\right)\cdot Q^{\Pi}\left(x_{i}\right)\right).
    \end{aligned}
\end{equation}
\end{definition}
The interpretation of $R^{\Pi}$ is the average welfare loss that the social planner expects from stochastic implementation of $g$ in $\mathscr{G}$ in the target population when $g$ is distributed according to $\Pi$.

We reiterate that stochastic assignment is achieved by randomly drawing $g$ according to $\Pi$. 
This way of selecting assignment rules is reminiscent of the \textit{Gibbs classifier} in statistical learning theory \citep{Germain2009} and might offer a computational advantage over other methods if drawing $g$ according to $\Pi$ is easier than maximising empirical welfare or finding the mode of $\Pi$, say.
An advantage of stochastic assignment is the possibility for sequential treatment evaluation:
the induced assignment of treatment and non-treatment to individuals in the target population by $\Pi$ is random conditional on $X$, which allows for estimation of the causal effect of treatment in future studies.
In this sense, stochastic assignment is well suited to balancing existing evidence about what constitutes the optimal assignment for each individual against the benefit of further exploration of the treatment effect \citep{manski2000using}. 
We do not, however, study this channel and focus on a purely static problem in this paper.

Our framework allows the social planner to hold some prior as to what constitutes the best policy.
We differentiate the subjective beliefs that the social planner holds, which we encode using the prior distribution $\pi_{0}$, and their updated beliefs following their observation of sample data, which we encode using $\Pi$. 
In the analysis that follows, we provide finite sample regret guarantees in the form of PAC-Bayes bounds for stochastic assignment. 
The approach discussed here differs significantly from other Bayesian treatment choice settings, such as those discussed in \citet{Chamberlain2011}, because we do not impose any kind of prior belief on $P$.
We instead choose to model the beliefs that the social planner has regarding the optimal policy. 
Using a decision procedure that is free from specification of the likelihood comes with a desirable robustness property, as we discuss in due course.

Aside from its more conventional role as a means of expressing existing information about what constitutes the best policy, $\pi_{0}$ can also play several other roles within our framework.
For instance, $\pi_{0}$ can also embed any constraints that are imposed upon the set of policies through truncation of its support.
Such a zero density condition can be imposed in lieu of an explicit restriction on $\mathscr{G}$ and is easy to implement in practice via rejection sampling, with only those policies that satisfy any budgetary, ethical or legal constraints being retained under the sampling procedure.
Moreover, $\pi_{0}$ can also be used to describe the status quo, with restrictions on the shape of $\pi_{0}$ governing how much policy can deviate.
These interpretations of $\pi_{0}$ naturally extend to $\Pi$.


\section{Optimal stochastic assignment and convergence of welfare}
\label{SEC:OPTIMALITY}


\subsection{Bounding expected welfare risk}
\label{SEC:DEVIATIONS}

Seeing as experimental data provides only an insight into the welfare performance of any policy in the target population, a very natural question to ask is how much we can expect $R^{\Pi}_{S}$ to differ from $R^{\Pi}$ for any given $\Pi$. 
We provide an answer to this question here.
\begin{theorem}
\label{THM:PAC} 
Suppose that \cref{ASS:EV,ASS:U,ASS:BO,ASS:SO} are satisfied, and $n\geq8$. 
Then, for any $0<\epsilon<1$ and $\Pi\in\mathscr{M}$ that is absolutely continuous with respect to $\pi_0$, the following inequality holds with probability at least $1- \epsilon$ in terms of $P^{n}$:
\begin{equation}
\label{EQ:PAC-BOUND}
    R^{\Pi} \leq R_{S}^{\Pi}+\sqrt{\frac{1}{2 n}\cdot\left(\mathrm{KL}\left(\Pi \| \pi_0\right)+\ln \left(\frac{2 \sqrt{n}}{\epsilon}\right)\right)}.
\end{equation}
\end{theorem}
We present a proof of this theorem in \cref{PROOF:PAC}. 
The proof builds upon \citet[\S Lemma 3]{Begin2016}, which offers a flexible approach that allows for the recovery of many different PAC-Bayes bounds.
The approach centres around a convex function of $R^{\Pi}$ and $R^{\Pi}_{S}$, which we specify so as to recover the form that is presented in \citet[see \S (8) in \citealp{Begin2016}]{McAllester2003}. 
We leverage results in \citet[\S Lemma 3 and \S Theorem 1]{Maurer2004}, exploiting the properties of Bernoulli random variables and convex functions, to adapt \citet{Begin2016} and the bound that is presented therein to allow for heterogeneous cost (in lieu of the standard binary loss function that is prevalent in the classification literature and that is studied in \citealp{Begin2016}).

We note that \cref{THM:PAC} holds for any rules that update $\pi_{0}$ and deliver $\Pi$, and that we have not committed to any particular updating rule to obtain \cref{EQ:PAC-BOUND}. 
A feature of \cref{EQ:PAC-BOUND} is that the regularisation term -- the square root term containing the Kullback-Leibler divergence of $\Pi$ from $\pi_{0}$ -- enters additively, which we find is a convenient feature for establishing convergence of our variational approximation (regularisation can otherwise be effected multiplicatively).\footnote{%
See \citet{Begin2016} for the implications of different convex functions.}
The regularisation term, by design, prevents overfitting.
To illustrate this point, suppose that $\pi_{0}$ is uniform over $\mathscr{G}$:
the best response of the social planner absent regularisation is to concentrate probability mass on the optimal (in-sample) $g$ as suggested by data, such that $\Pi$ is degenerate.
In the presence of regularisation, however, this is no longer a best response, since the Kullback-Leibler divergence infinitely penalises degeneracy \textit{vis-\`{a}-vis} uniformity.
Rather, the best response of the social planner is to allocate probability mass on all $g$ in $\mathscr{G}$, albeit concentrating more mass on those $g$ that are associated with low empirical welfare risk.
Put differently, the regularisation term controls how far away from $\pi_{0}$ a stochastic assignment rule can be, with this difference governed by the number of observations in the sample.

The Vapnik-Chervonenkis (VC) dimension is the standard measure of complexity in the statistical learning literature.
We instead associate complexity with the Kullback-Leibler divergence. 
If one is willing to impose distributional constraints on a posterior assignment rule, an advantage of the PAC-Bayes approach and of using the Kullback-Leibler divergence is that complexity is then purely in terms of the selected (stochastic) assignment rule $\Pi$ rather than in terms of the class of possible stochastic assignment rules $\mathscr{M}$ or the class of underlying deterministic rules $\mathscr{G}$.
As such, \cref{THM:OPTIMAL} does not explictly require any assumption about the VC dimension of $\mathscr{G}$.
The influence of VC dimension for $\mathscr{G}$ is implicit in our setting:
the upper bound on the difference between $R_{S}^{\Pi}$ and $R^{\Pi}$ implied by \cref{EQ:PAC-BOUND} is governed by the Kullback-Leibler divergence, which is increasing in the dimension of the support of $\Pi$ and $\pi_{0}$, and so is non-decreasing in the complexity of $\mathscr{G}$.


\subsection{Optimal updating rule}
\label{SEC:UPDATING}

In a standard Bayesian setting, unknown parameters index the distribution of data and inference on parameters is conducted with respect to the posterior distribution.
Typically, the posterior distribution is constructed from a well-defined likelihood function via Bayes' theorem.
We leverage \cref{THM:PAC} to construct $\Pi$ from $\pi_{0}$ and $R^{\Pi}_{S}$.
This approach is valid since \cref{THM:PAC} holds for all $\Pi\in\mathscr{M}$.
We emphasise that the posterior distribution that we construct is over assignment rules rather than over the data generating processes, which is distinct from \citet{Bissiri2016} and \citet{csaba2020learning}.

Following \citet{McAllester2003} and \citet{Germain2009}, we define an optimal posterior distribution, which we denote by $\Pi^{*}$, as a distribution over $\mathscr{G}$ that minimises the right hand side of \cref{EQ:PAC-BOUND}. That is, $\Pi^{*}$ minimises
\begin{equation}
     R_{S}^{\Pi} +\sqrt{\frac{1}{2 n}\cdot\left( \mathrm{KL}\left(\Pi \| \pi_0\right) \text+\ln\left(\frac{2 \sqrt{n}}{\epsilon}\right)\right)}.
\end{equation}
\begin{theorem}
\label{THM:OPTIMAL} 
The optimal posterior $\Pi^{\ast}$ over $\mathscr{G}$ satisfies
\begin{equation}
    \mathrm{d}\Pi^{*}\left(g\right)
    \equiv
    \frac{\exp \left(-\chi\cdot R_{S}\left(g\right)\right)}{\int_{\mathscr{G}} \exp \left(-\chi\cdot R_{S}\left(g\right)\right)\cdot \mathrm{d}\pi_{0}\left(g\right)} \cdot \mathrm{d}\pi_{0}\left(g\right),
\end{equation}
where
\begin{equation} 
\label{EQ:OPTIMAL-CHI}
\chi 
\equiv 
4 n\cdot\sqrt{\frac{1}{2 n}\cdot\left(\mathrm{KL}\left(\Pi^* \| \pi_0\right)+\ln\left( \frac{2 \sqrt{n}}{\epsilon}\right)\right)}.
\end{equation}
\end{theorem}
We present a proof of this theorem in \cref{PROOF:OPTIMAL}.

The posterior distribution that we derive is analogous to the optimal posterior in \citet{McAllester2003} with the difference that our observations are mapped to the unit interval rather than to $\{-1,1\}$, which is the standard support in classification.
This particular distribution is common in the statistical mechanics literature and is a Boltzmann (or Gibbs) distribution, and has the form of exponential tilting of the prior, where the exponential tilting term involves the negative empirical welfare risk. 
The degree of tilting depends upon the magnitude of $\chi>0$, which is the inverse of the Lagrange multiplier of the associated minimisation problem and that corresponds to the root of \cref{EQ:OPTIMAL-CHI}. 
The Lagrange multiplier controls the extent to which $\pi_{0}$ is updated by empirical welfare risk in minimising the upper bound for $R^{\Pi}$.
 
Although \cref{THM:OPTIMAL} offers an analytical characterisation of the optimal posterior, we are unable to obtain a closed-form expression for the optimal posterior density. 
Whilst \cref{EQ:OPTIMAL-CHI} does suggest a means to compute this density, it is likely that, in practice, this computation is difficult to perform with any precision.
Given that we have, however, established that this density exists, we can consider approximating it using the \emph{variational approximation} method, as is considered in \citet{Alquier2016}.

 
\subsection{Variational approximation of the optimal stochastic assignment rule}
\label{SEC:APPROXIMATION}

We develop a variational approximation of the optimal posterior density. 
Variational approximation is useful in situations where Gibbs distributions are difficult to sample from directly,
such as is the case for graphical models where Markov Chain Monte Carlo (MCMC) sampling is costly \citep{Wainwright2008}.
In variational approximation, we choose to approximate the optimal posterior distribution via a family of distributions of our choice, $\mathscr{V}\subset\mathscr{M}$.
This allows us to develop an analytically tractable upper bound for the welfare regret attained by the resulting stochastic assignment rule. 

Aside from guaranteeing tractability in estimation, variational approximation can also be motivated as a convenient way to impose constraints on the set of policies.
For instance, a fairness criterion requiring that individuals with similar characteristics have similar probabilities of treatment \citep{dwork2012fairness} can be enforced by specifying that $\mathscr{V}$ is a continuous family, and by limiting the concentration of the density.
A similar approach can be used if $\mathscr{V}$ is a parametric family to limit how much policy can deviate from the status quo, by fixing the parameters of the posterior distribution or restricting them to some set, say.

We approximate $\Pi^{\ast}$ (implicitly defined in \cref{THM:OPTIMAL}) by minimising the right-hand side of \cref{EQ:PAC-BOUND} with respect to posterior distributions in $\mathscr{V}$, defining
\begin{equation}
    \tilde{\Pi}
    \equiv
    \underset{\Pi \in \mathscr{V}}{\operatorname{argmin}} \,\left\lbrace R_{S}^{\Pi} + \sqrt{\frac{1}{2 n}\cdot\left(\mathrm{KL}\left(\Pi \| \pi_{0}\right)+\ln \left(\frac{2 \sqrt{n}}{\epsilon}\right)\right)}\right\rbrace.
\end{equation}
We then use this optimal variational posterior distribution to define a new bound for welfare regret from which we can characterise its convergence rate. 
\begin{lemma}
\label{THM:VARIATIONAL} 
With probability $1-\epsilon$ in terms of $P^n$, the expected welfare risk under the optimal variational posterior satisfies 
\begin{equation}
    R^{\tilde{\Pi}} \leq \inf _{\Pi \in \mathscr{V}} \left\{ R^{\Pi} +\frac{a\left(\lambda, n\right)}{\lambda}+\frac{\mathrm{KL}\left(\Pi \| \pi_0\right)}{\lambda}+\frac{\ln \left(\frac{2}{\epsilon}\right)}{\lambda}+\sqrt{\frac{1}{2 n}\cdot\left(\mathrm{KL}\left(\Pi \| \pi_{0}\right)+\ln \left(\frac{4 \sqrt{n}}{\epsilon}\right)\right)} \right\},
\end{equation}
where $\epsilon\in\left(0,1\right)$ and $\lambda > 0$ are arbitrary constants and $a$ is a function that depends upon a positive constant $\lambda$ and the sample size $n$.
\end{lemma}
In what follows, we let $\mathscr{V}$ be a variational family of distributions that assigns positive density to assignment rules in $\mathscr{F}$ only.
Equivalently, we let $\mathscr{V}$ be a directional family that assigns positive density to unit vectors on the hypersphere, recalling that every policy in the LES class can be uniquely associated with a unit vector (see \cref{EQ:UNIQUENESSOFINDEX} and surrounding discussion).
A particularly tractable directional family, and one that we use, is the von Mises-Fisher family of distributions, which is characterised by a probability density function satisfying, for all $\kappa>0$ and $m$-dimensional unit vectors $\mu$,
\begin{equation}
    \mathrm{d}\Pi\left(\beta;\kappa,\mu\right)
    \equiv
    \frac{\kappa^{m/2-1}\cdot \exp\left({\kappa\cdot\mu^{\top}\beta}\right)}{\left(2\pi\right)^{m/2}\cdot I_{m/2-1}\left(\kappa\right)}\cdot\mathrm{d}\beta,
\end{equation}
where $I_{\nu}\left(z\right)$ is a modified Bessel function of the first kind with order $\nu$ and argument $z>0$.
A von Mises-Fisher distribution is the analogue of a multivariate Gaussian distribution on the unit hypersphere.
We refer to $\kappa$ as the concentration parameter and to $\mu$ as the mean direction (the location parameter), noting that the distribution becomes degenerate as $\kappa\rightarrow\infty$ and uniform as $\kappa\rightarrow 0$.

We now introduce some further notation that facilitates our analysis. 
First, we let $\overline{R} = \inf_{\Pi \in \mathscr{M}} R^{\Pi}$ denote the minimum expected welfare risk amongst assignment rules in $\mathscr{F}$ and $\overline{\beta}$ denote the vector that parametrises the assignment rule that induces $\overline{R}$. 
We also add the following assumption that restricts the marginal distribution of $X$.
\begin{assumption}[Margin assumption]
\label{ASS:CO}
There exists a constant $c>0$ such that, for any $m$-dimensional unit vectors $\beta^{\dagger}$ and $\beta^{\ddagger}$, $P\left(\langle X, \beta^{\dagger}\rangle\cdot\left\langle X, \beta^{\ddagger}\right\rangle \leq 0\right) 
\leq 
c\cdot\left\|\beta^{\dagger}-\beta^{\ddagger}\right\|_2$.
\end{assumption}
This assumption is satisfied whenever $X$ has bounded density on the unit hypersphere and is also present in the analysis of \citet{Alquier2016}. 
An interpretation of this assumption is that the proportion of individuals whose treatment status switches is continuous with respect to the linear eligibility score coefficients, with an implication being that $R\left(\beta\right)-\overline{R}\leq 2c\cdot\|\beta - \overline{\beta}\|_2$.

We now present a high-probability uniform upper bound for the welfare regret of the stochastic assignment rule obtained by variational approximation via the von Mises-Fisher family of distributions.
\begin{theorem}
\label{THM:RATE}
Suppose that \cref{ASS:EV,ASS:U,ASS:BO,ASS:SO,ASS:CO} are satisfied, that $\pi_0$ is a uniform distribution over the unit hypersphere, and that $\Pi$ is a von Mises-Fisher distribution. 
Then for $n\geq8$,
with probability at least $1-\epsilon$ in terms of $P^n$,
\begin{equation}
\label{EQ:DECAY}
    R^{\tilde{\Pi}} - \overline{R} 
    \leq 
    M\cdot\frac{\ln\left(n\right)}{\sqrt{n}},
\end{equation}
where $M$ is a universal constant.
\end{theorem}
We present proof of this theorem and provide an analytical expression for the universal constant in \cref{PROOF:RATE}.

The uniform upper bound on welfare regret that is defined by \cref{EQ:DECAY} decays at a rate of  $\ln\left(n\right)/\sqrt{n}$.
This rate is slightly slower than the welfare regret convergence rate of the EWM (deterministic) assignment rule studied in \citet{Kitagawa2018a}. 
A simple comparison of these rates, however, is not quite meaningful for the following reason:
we do not know if the convergence rate of $\ln\left(n\right)/\sqrt{n}$ that is obtained in \cref{THM:RATE} is sharp or not. 
\cref{THM:RATE} requires \cref{ASS:CO}, whilst \citet{Kitagawa2018a} does not impose this assumption in showing that $1/\sqrt{n}$ is the minimax optimal rate of welfare regret convergence.\footnote{%
\citet[\S Theorem 2.3 and \S Theorem 2.4]{Kitagawa2018a} establishes that the minimax-optimal rate under a stronger condition than our \cref{ASS:CO} is $1/n^{2/3}$.
This stronger condition embeds a margin assumption that implies our \cref{ASS:CO} but also embeds the requirement that the first-best treatment rule is contained in the set of admissible decision rules -- that $\mathscr{F}$ contains the deterministic assignment rule that minimises expected welfare risk amongst all assignment rules.
We do not make any assumption about whether $\mathscr{F}$ (or indeed $\mathscr{G}$ in earlier parts of our analysis) contains the first-best assignment rule, or whether this rule is deterministic or stochastic.} 
We do not know what the minimax optimal rate of welfare regret convergence is when \cref{ASS:CO} is additionally imposed and, hence, cannot rule out the possibility that the convergence rate of \cref{THM:RATE} can be improved upon and made faster than $1/\sqrt{n}$.
We leave further investigation of this matter for future research.

It is worth emphasising that the regret convergence result of \cref{THM:RATE} imposes weak restrictions on the distribution of data (Assumptions \ref{ASS:BO} - \ref{ASS:CO}) and does not require the specification of a likelihood function or of regression equations. 
This contrasts with other approaches such as a Bayesian approach, where misspecification of likelihood can lead to non-convergence of the welfare regret even when Bayes optimal policies are constrained to deterministic ones in $\mathscr{F}$. 

Our approach is similar to \citet{Alquier2016} in which the families of distributions for variational approximation are multivariate Gaussian distributions on Euclidean space with flexible covariance matrices.
In our approach, we stipulate a class of von Mises-Fisher distributions, which are the hyperspherical analogues of multivariate Gaussian distribution with diagonal covariance matrices featuring a constant variance element. 
Since the empirical welfare criterion for LES rules is invariant to the scale of $\beta$, it is natural to consider von Mises-Fisher distributions rather than Gaussian ones. The scale invariance of von Mises-Fisher distributions can simplify optimisation of the variational approximation by reducing the set of optima to a singleton. 


\section{Implementation}
\label{SEC:IMPLEMENTATION}

To implement our procedure, we restrict attention to $\mathscr{F}$ and let $\mathscr{V}$ be the von Mises-Fisher family of distributions.
Our goal is then to minimise the objective function,
\begin{equation}
\label{IMPLEMENTATION:OBJECTIVE}
    R^{\Pi}_S+\sqrt{\frac{1}{2n}\cdot\left(\text{KL}\left(\Pi|\pi_0\right)+\ln\left(\frac{2\sqrt{n}}{\epsilon}\right)\right)},
\end{equation}
with respect to $\kappa$ and $\mu$, which are the parameters of our chosen variational family (the concentration parameter and the mean direction, respectively). 
Here, we reiterate that $\epsilon$ relates to the probability with which our high probability bounds hold.
We assume that $\pi_0$ is the uniform distribution over the sphere and set $\epsilon$ equal to the 5\% level throughout.

We propose numerically minimising the objective function, approximating $R^{\Pi}_{S}$ using Monte Carlo draws of $\beta$ from the von Mises-Fisher distribution for a given realisation of data and for fixed values of the parameters of the von Mises-Fisher distribution.
We let $\{\beta^j : j=1,...,J\}$ be the pseudo-random draws that we obtain, and compute 
\begin{equation}
\label{EQ:RISK-IMPLEMENTATION}
    \hat{R}^{\Pi}_{S}
    \equiv
    \frac{1}{n}\sum_{i=1}^{n}h_{i}\cdot\left(D_{i}\cdot\left(1-\hat{\Pi}_{i}\right)+\left(1-D_i\right)\cdot\hat{\Pi}_{i}\right),
\end{equation}
where, for all $i=1,...,n$,
\begin{equation}
    \hat{\Pi}_{i}\equiv
    \frac{1}{J}\sum_{j=1}^{J}1\left(x_{i}^{\top}\beta^j\geq 0\right).
\end{equation}
Fast pseudo-random sampling of von Mises-Fisher random vectors is possible using the rejection sampling method of \citet{wood1994simulation} or the inversion method of \citet{kurz2015stochastic}.
The analogue of \cref{IMPLEMENTATION:OBJECTIVE} that we then minimise is
\begin{equation}
\label{IMPLEMENTATION:OBJECTIVE2}
\hat{R}^{\Pi}_S+\sqrt{\frac{1}{2n}\cdot\left(
\nu\cdot\ln\left(\frac{\kappa}{2}\right)-\ln\left(I_{\nu}\left(\kappa\right)\right)-\ln\left(\Gamma\left(\nu+1\right)\right)+\frac{I_{\nu+1}\left(\kappa\right)}{I_{\nu}\left(\kappa\right)}\cdot\kappa
+\ln\left(\frac{2\sqrt{n}}{\epsilon}\right)\right)},
\end{equation}
which follows from our assumption of a uniform prior and the form that the Kullback-Leibler divergence takes under that assumption \citep{KitagawaRowley}.

As is shown in \citet{KitagawaRowley}, the Kullback-Leibler divergence of the von Mises-Fisher distribution from the uniform distribution over the hypersphere does not depend upon $\mu$ and increases at the logarithmic rate in $\kappa$. 
In contrast, although $R^{\Pi}_{S}$ is a function of $\mu$ and $\kappa$, the nature of this relationship is not clear. 
That several combinations of $\mu$ and $\kappa$ induce local minima of $R^{\Pi}_{S}$ cannot be ruled out for instance.
Moreover, since we construct $\hat{R}^{\Pi}_{S}$ based upon random draws of $\beta$ from $\Pi$, our objective function is not a smooth function of $\mu$ and $\kappa$. 
Given this, we suggest that a grid-based search for the minimum of the objective function is appropriate, with this search limited to values of $\kappa$ between zero and some specified upper limit. 
Further insight about the behaviour of the objective function is provided in an online appendix.


\section{Empirical illustration}
\label{SEC:ILLUSTRATION}

We illustrate our procedure using data from the National Job Training Partnership Act (JTPA) Study.
Applicants to the Study were randomly allocated to one of two groups.
Applicants allocated to the treatment group were extended training, job search assistance and other services provided by the JTPA over a period of 18 months.
Applicants allocated to the control group were excluded from JTPA assistance.
Along with information collected prior to the commencement of the intervention, the Study also collected administrative and survey data relating to applicants' earnings in the 30 months following its start. 
Further details about the data and the Study can be found elsewhere \citep[see, for instance,][]{bloom1997benefits}.
We restrict attention to a sample of 9,223 observations for which data on years of education and pre-programme earnings amongst the sample of adults (aged 22 years and older) used in the original evaluation of the programme and in subsequent studies (\citealp{bloom1997benefits,heckman1997matching,abadie2002instrumental}) is available.
Applicants in this sample were assigned to the treatment group with a probability of two thirds. 
Like \citet{Kitagawa2018a}, we define $D$ to be the initial assignment of treatment, rather than the actual take-up due to the presence of non-compliance in the experiment.
We study stochastic assignment rules.

We follow \citet{Kitagawa2018a} in considering total individual earnings in the 30 months after programme assignment as our principal welfare measure.
Moreover, we focus exclusively on the class of linear rules,
\begin{equation}
\label{EMPIRICS:FCLASS}
\begin{aligned}
    \mathscr{F}
    &=
    \left\{x : \beta_{0}+\beta_{1}\cdot\text{prior earnings}+\beta_{2}\cdot\text{education}\geq 0\left\vert\,\|\left(\beta_{0},\beta_{1},\beta_{2}\right)\|_{2}=1\right.\right\},\\
    &=
    \left\{x : \beta_{0}+\beta_{1}\cdot x_{1}+\beta_{2}\cdot x_{2}\geq 0\left\vert\,\|\left(\beta_{0},\beta_{1},\beta_{2}\right)\|_{2}=1\right.\right\},\\
\end{aligned}
\end{equation}
that are studied in that paper.

\begin{figure}
    \centering
    \caption{Variation in treatment propensity across individuals in the JTPA Study sample}
    \label{FIG:PS}
    \includegraphics[width=\textwidth,height=3.5in]{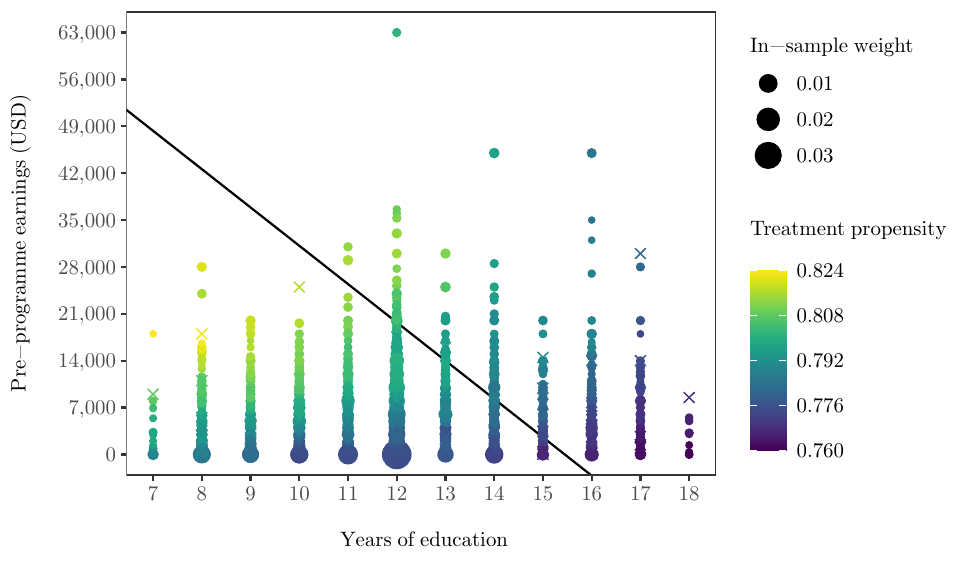}
    \caption*{\footnotesize\textit{Notes:} 
    JTPA Study sample.
    This figure illustrates the treatment propensity of individuals under the posterior assignment rule that is induced by $\left(\kappa^{*},\mu^{*}\right)$.
    Each point represents the individual characteristics of an individual or several individuals in the (crosses denote individuals with zero in-sample weight).
    For comparison, individuals to the left of the solid diagonal line are assigned treatment under the optimal deterministic assignment rule of \citet{Kitagawa2018a}.}
\end{figure}

To implement our procedure, we map prior earnings and education to the unit interval,\footnote{%
We map each variable to the unit interval by dividing through by its maximum in the sample.
\citet{Kitagawa2018a} also does this.
Such a change of units is useful when the domain of one variable is much larger than the domain of another and the respective coefficients on the two variables reflect this.
For instance, in our sample, every individual has between seven and 18 years of education, and no individual earned more than \$63,000 prior to the start of the intervention.} 
and calculate $H$ as outlined in \cref{EQ:WEIGHTSFORIMPLEMENTATION}.
We perform this calculation without adjusting post-programme earnings by the average cost of JTPA assistance (\$774 per individual) for treated individuals.\footnote{%
We adjust post-programme earnings by the average cost of JTPA assistance in an online appendix.}  
We then utilise a grid search approach over the parameters of the von Mises-Fisher distribution, specifying a reasonably fine grid over the unit sphere and over a finite subset of the reals.\footnote{%
We design our grid so as to place an upper limit on the great-circle distance between any point on the sphere and its closest point on the grid.
Our grid comprises a total of 10,116 directional vectors combined with a sequence of evenly-spaced concentrations on the $[0,5]$ interval.
For reference, the surface area of the sphere is $4\pi$, which means that our grid has a density of approximately $0.001$.} 

For each point on our grid, we draw 1,000 values of $\beta$ from the corresponding von Mises-Fisher distribution and approximate empirical welfare risk as per \cref{EQ:RISK-IMPLEMENTATION}.
We then substitute these values into \cref{IMPLEMENTATION:OBJECTIVE2} to provide an estimate of the objective function.\footnote{%
In \cref{IMPLEMENTATION:OBJECTIVE2}, given \cref{EMPIRICS:FCLASS} and its restriction of $\beta$ to the unit sphere, $\nu=1/2$ .
More generally, maintaining our convention of defining $\mathscr{X}\in\mathbb{R}^{m-1}$ and adding to this vector a constant (i.e., an intercept), $\nu=m/2-1$.}

We find that the objective function is minimised (amongst the class of von Mises-Fisher distributed linear assignment rules) by the stochastic assignment rule with $\kappa=1.550$ and $\mu=\left(0.883,0.442,0.158\right)$,\footnote{%
This directional vector can be represented by an azimuth of $27^\circ$ and an inclination of $81^\circ$ using spherical coordinates.}
which we label as $\kappa^{*}$ and $\mu^{*}$, respectively.

\begin{figure}
    \centering
    \caption{Behaviour of the objective function at $\mu^{*}$ given variation in $\kappa$}
    \label{FIG:KAPPA}
    \includegraphics[width=\textwidth,height=3.5in]{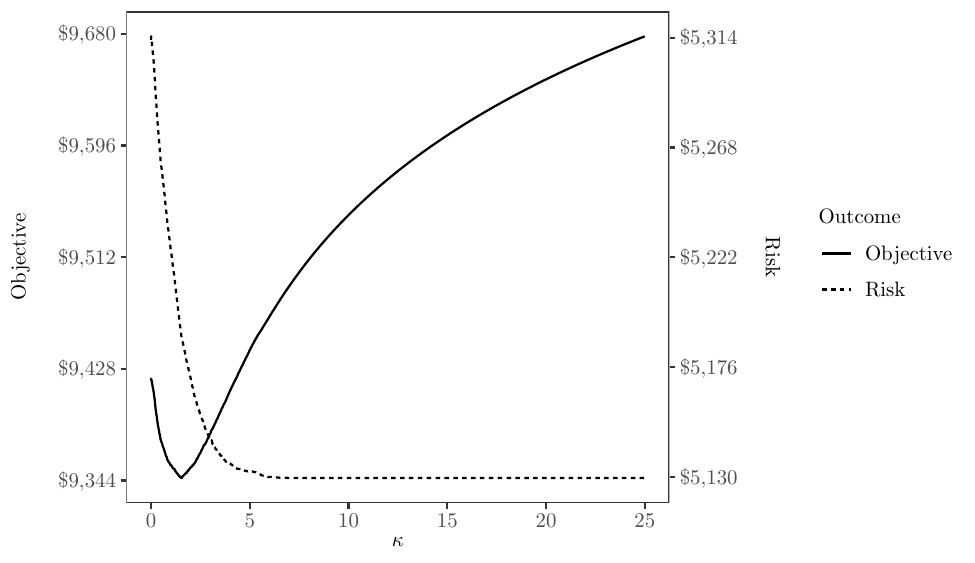}
     \caption*{\footnotesize\textit{Notes:} 
    JTPA Study sample.
    This figure illustrates the shape of the objective function and its risk component at $\mu^{*}$ as $\kappa$ is varied;
    high concentration is associated with low risk but incurs a large penalty for divergence from the uniform prior. }
\end{figure}

The optimal stochastic assignment rule, on average, assigns treatment to individuals in the JTPA Study sample around 83\% of the time.
This probability is not, however, uniform, and there is some variation in the probability with which distinct individuals are assigned treatment.
This variation in assignment propensity can be seen in \cref{FIG:PS}, which plots the individual characteristics of all individuals in the sample.
The propensity with which individuals with distinct characteristics are assigned treatment is represented by the color of each point, and the weight given to individuals in the sample with particular characteristics is represented by the size of each point.
The weight attached to a given point is proportional to the sum of post-programme earnings over all individuals with those characteristics.\footnote{%
To simplify \cref{FIG:PS}, we scale the weights such that they sum to one.}
\cref{FIG:PS} shows that the optimal stochastic assignment rule is more likely to assign treatment to an individual with few years of education and high prior earnings than an individual with more years of education and lower prior earnings, with the assignment probability ranging from 78\% to 84\%. 
The deterministic assignment rule of \citet{Kitagawa2018a}, in contrast, assigns only individuals with few years of education and low prior earnings to treatment, with around 93\% of individuals assigned treatment.
We plot this deterministic rule as a useful benchmark for comparison in \cref{FIG:PS}.

It is important to emphasise that it is the regularisation term and, in particular, the Kullback-Leibler divergence that limits the value of $\kappa$ at the optimum, and leads to an interior probability of assignment for all individuals.
This can be seen in \cref{FIG:KAPPA}, which plots the USD equivalent of the objective function (left-hand axis, solid line) and of empirical welfare risk (right-hand axis, dashed line) for a range of $\kappa$, holding fixed $\mu$ at $\mu^{*}$.
We observe that empirical welfare risk decreases as $\kappa$ increases, remaining low and constant once its value is sufficiently large.\footnote{%
That empirical welfare risk decreases for small to moderate values of $\kappa$ is specific to the data and chosen $\mu$, and is arguably also attributable to the lack of consideration given to the cost of treatment.}
The intuition here is that large values of $\kappa$ lead to stochastic assignment rules that mimic deterministic ones;
we expect $\mu$ to eventually coincide with the deterministic assignment rule of \citet{Kitagawa2018a} as the value of the concentration parameter approaches infinity,since there does not exist a (linear) deterministic rule that can improve upon this.  
Tempering this preference towards large values of $\kappa$ is the Kullback-Leibler divergence of the von Mises-Fisher distribution from the uniform distribution, which is increasing at the logarithmic rate in $\kappa$ and generates the difference between the objective function and empirical welfare risk in \cref{FIG:KAPPA}.
As $\kappa$ increases, the regularisation term begins to dominate.

\begin{figure}
    \centering
    \caption{Deterministic assignment rules and empirical welfare risk}
    \label{FIG:DETERMINISTICRULES}
    \includegraphics[width=\textwidth,height=3.5in]{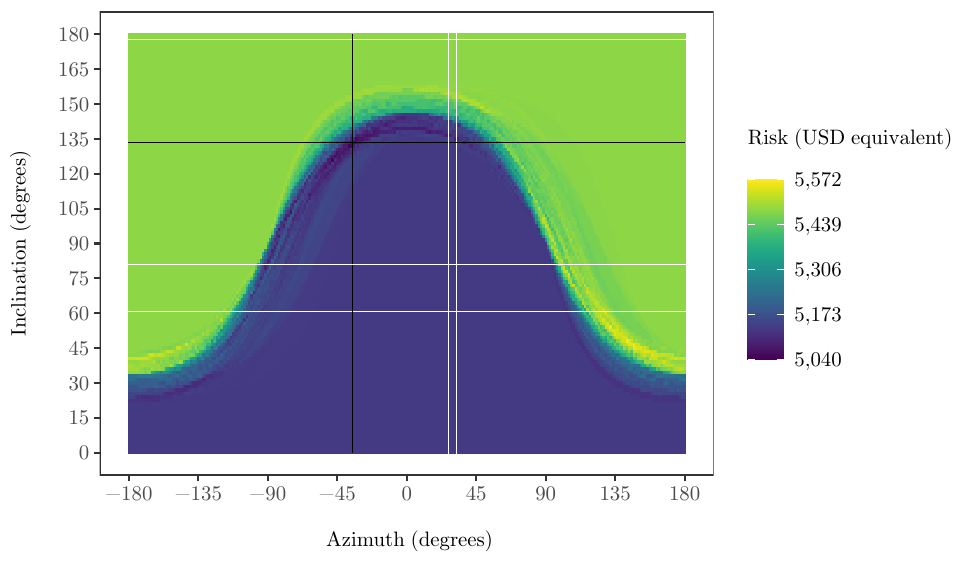}
    \caption*{\footnotesize\textit{Notes:} 
    JTPA Study sample.
    This figure illustrates the risk that is associated with (deterministic) assignment rules in $\mathscr{F}$.
    A spherical coordinate mapping is implemented.
    The intersection of the two white lines is located at $\mu^{*}$. 
    The intersection of the two black lines is located at the optimal deterministic assignment rule of \citet{Kitagawa2018a}, which attains the minimal regret amongst all deterministic linear rules.}
\end{figure}

\begin{figure}
    \centering
    \caption{Behaviour of the objective function at $\kappa^{*}$ given variation in the mean direction $\mu$}
    \label{FIG:MU}
    \includegraphics[width=\textwidth,height=3.5in]{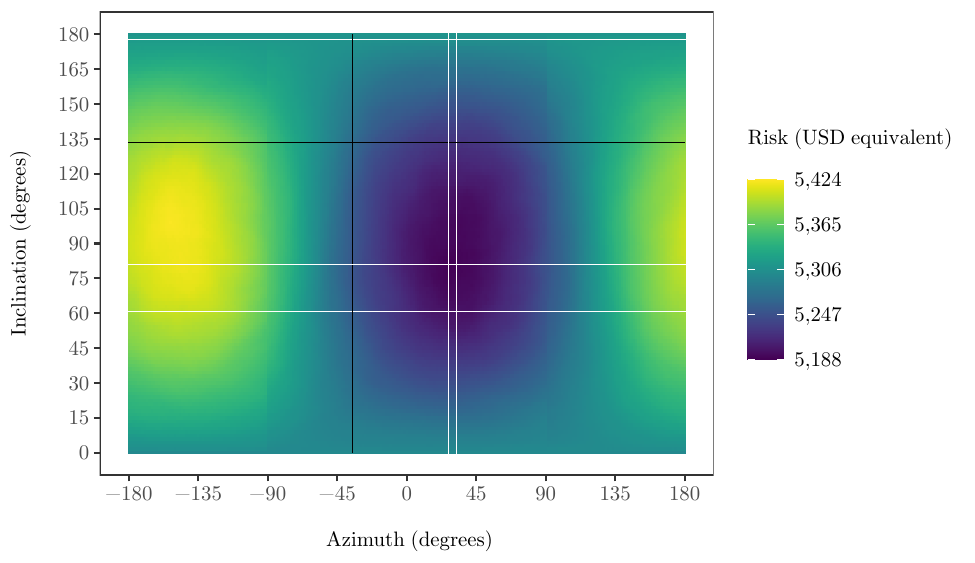}
    \caption*{\footnotesize\textit{Notes:} 
    JTPA Study sample.
    This figure illustrates the risk that is associated with (stochastic) assignment rules in $\mathscr{V}$;
    the concentration parameter is fixed at $\kappa^{*}$ whilst $\mu$ is varied.
    The intersection of the two white lines is located at $\mu^{*}$. 
    The intersection of the two black lines is located at the optimal deterministic assignment rule of \citet{Kitagawa2018a}, which attains the minimal regret amongst all deterministic linear rules.}
\end{figure}

To better understand why $\mu^*$ does not coincide with the deterministic assignment rule of \citet{Kitagawa2018a} holding fixed $\kappa$ at $\kappa^*$, we refer to \cref{FIG:DETERMINISTICRULES}, which plots empirical welfare risk for all vectors on the unit sphere -- i.e., the empirical welfare risk associated for each deterministic assignment rule in $\mathscr{F}$.
\cref{FIG:DETERMINISTICRULES} utilises the spherical coordinate system
\begin{equation}
    \left(\beta_0,\beta_1,\beta_2\right)=\left(\cos\left(\theta\right)\sin\left(\phi\right),\sin\left(\theta\right)\sin\left(\phi\right),\cos\left(\phi\right)\right),
\end{equation}
where $\theta\in\left[-180^\circ,180^\circ\right)$ is the azimuth and $\phi\in\left[0,180^\circ\right]$ is the inclination. 
It is perhaps convenient to think of the azimuth as related to longitude and the inclination as related to latitude.
For non-trivial $\kappa$, the von Mises-Fisher distribution allocates probability mass to the sphere in such a way that its density contours are concentric about $\kappa$, with points closer to $\mu$ more likely to occur.
The deterministic assignment rule of \citet{Kitagawa2018a} can be seen from \cref{FIG:DETERMINISTICRULES} to be located on the boundary between a high risk region (no-one treated) and a moderate risk region (everyone treated).
As such, a stochastic assignment rule with non-trivial $\kappa$ and $\mu$ located at this point would approximately allocate probability mass to each of these regions in equal amounts.
By shifting $\mu$ towards the centre of the moderate risk region, we allocate relatively more mass to rules that induce moderate risk and less mass to rules that induce high risk, which reduces empirical welfare risk overall.

This pattern underlies what we observe in \cref{FIG:MU}, which plots empirical welfare risk for all $\mu$ on the sphere holding fixed $\kappa$ at $\kappa^{*}$.
Despite the apparent discontinuity of risk over deterministic assignment rules, empirical welfare risk (and the objective function) appear to vary smoothly. 


\section{Conclusion}
\label{SEC:CONCLUSION}

Central to our analysis of the treatment choice problem is the question of \textit{how should the social planner allocate individuals to treatment with a given probability rather than with certainty?}
To answer this question, we focus on stochastic assignment rules that we formulate as posterior distributions obtained from well-defined prior distributions via a PAC-Bayes approach.
These distributions are able to accommodate any initial belief that the social planner holds about what constitutes the best treatment, as well as any budgetary, ethical or legal constraints that are imposed and that can arise due to concerns about fairness or about maintaining the status quo.
We establish that it is possible to obtain the minimum expected welfare risk under a variational approximation of the optimal posterior distribution, which we also characterise.
In keeping with the notion that variational approximation replaces a general class of assignment rules with a simpler set of policies, we focus on stochastic assignment rules that can be expressed as density functions over the LES class of (deterministic) assignment rules. 
We exploit the scale invariance of the LES class to restrict attention to distributional families on the unit hypersphere, selecting the von Mises-Fisher family of distributions as our chosen variational family.
We demonstrate how our methods can be used in an empirical setting by estimating which individuals should be entered onto a job training programme, using data from the well-known JTPA Study.

Our research suggests several further questions that remain unanswered.
\textit{How does the choice of variational family affect the rate of convergence?}
The von Mises-Fisher family of distributions is not the only directional family -- other families such as the Matrix Langevin and Kent families allow for richer covariance structures.
Whilst more general distributions admit more complex stochastic assignment rules, does this additional generality come at an additional cost?
\textit{Do other parametrisations of the class of assignment rules exist that can achieve faster decay rates?}
We establish that the optimal stochastic assignment rule in our chosen variational family yields a convergence rate for welfare regret of $\ln\left(n\right)/\sqrt{n}$. 
Ignoring the logarithm in the numerator, this rate coincides with the minimax optimal rate for deterministic assignment rules in the absence of the margin assumption, as shown in \citet{Kitagawa2018a}.
Our convergence rate result, however, relies upon Assumption \ref{ASS:CO} that constrains the marginal distribution of $X$.
We do not know whether this assumption leads to the minimax optimal rate being faster than $1/\sqrt{n}$ and whether the proposed method can attain a faster convergence rate than $\ln\left(n\right)/\sqrt{n}$. 


\appendix


\section{Proofs of theorems and associated lemmata}
\label{SEC:PROOF}


\subsection{Proof of \cref{THM:PAC}}
\label{PROOF:PAC}

We let $c : \left[0,1\right]^{2}\rightarrow\mathbb{R}$ denote a convex function. 
Then
\begin{equation} 
\label{EQ:CONVEXITY}
\begin{aligned}
    n\cdot c\left(R^{\Pi}_{S},R^{\Pi}\right) 
    &= 
    n\cdot c\left[\mathrm{E}_{\Pi}\left(R_{S}\left(g\right)\right),\mathrm{E}_{\Pi}\left(R\left(g\right)\right)\right],\\
    &\leq 
    \mathrm{E}_{\Pi}\left[n\cdot c\left(R_{S}\left(g\right),R\left(g\right)\right)\right],
    \end{aligned}
\end{equation}
by convexity.
\begin{namedtheorem}[\textbf{\citet[\S Lemma 3]{Begin2016}}\textrm{(Kullback-Leibler change of measure)}] 
Let $\phi : \mathscr{G}\rightarrow\mathbb{R}$ be a measurable function. 
For any $g \in \mathscr{G}$ and any distributions $\Pi$ and $\pi_{0}$ on $\mathscr{G}$ such that $\Pi$ is absolutely continuous with respect to $\pi_{0}$,
\begin{equation}
    \mathrm{E}_{\Pi}\left( \phi\left(g\right)\right) 
    \leq 
    \mathrm{KL}\left(\Pi \| \pi_0\right)+\ln \left[\mathrm{E}_{\Pi}\left( \exp\left({\phi\left(g\right)}\right)\right)\right].
\end{equation}
\end{namedtheorem}
We apply the lemma above, with $\phi\left(g\right) =  n\cdot c\left(R_{S}\left(g\right),R\left(g\right)\right)$, to obtain
\begin{equation} 
\label{EQ:BEGIN-LEMMA-3}
     \mathrm{E}_{\Pi}\left[ n\cdot c\left(R_{S}\left(g\right),R\left(f\right)\right)\right]
     \leq 
     \mathrm{KL}\left(\Pi \| \pi_{0}\right)+\ln \left(\mathrm{E}_{\Pi}\left( \exp\left[n\cdot c\left(R_{S}\left(g\right),R\left(g\right)\right) \right]\right)\right).
\end{equation}
We then apply Markov's inequality to $\mathrm{E}_{\Pi}\left(\exp\left({n\cdot c\left(R_{S}\left(g\right),R\left(g\right)\right)}\right)\right)$ to obtain
\begin{equation}
    P^{n}\left( \mathrm{E}_{\Pi}\left(\exp\left({n\cdot c\left(R_{S}\left(g\right),R\left(g\right)\right)}\right)\right) \geq \frac{1}{\epsilon} \cdot\mathrm{E}_{P^{n}}\left(\mathrm{E}_{\Pi}\left( \exp\left[n\cdot c\left(R_{S}\left(g\right),R\left(g\right)\right)\right]\right) \right)\right)
    \leq 
    \epsilon.
\end{equation}
Thus, with probability at least $1-\epsilon$ we have, for all $\Pi \in \mathscr{M}$,
\begin{equation} 
\label{EQ:EXPECTATION-BOUND}
    \mathrm{E}_{\Pi}\left(\exp\left[n\cdot c\left(R_{S}\left(g\right),R\left(g\right)\right)\right]\right)
    \leq 
    \frac{1}{\epsilon}\cdot \mathrm{E}_{P^{n}}\left(\mathrm{E}_{\Pi}\left( \exp\left[n\cdot c\left(R_{S}\left(g\right),R\left(g\right)\right)\right]\right)\right).
\end{equation}
Combining \cref{EQ:CONVEXITY,EQ:BEGIN-LEMMA-3,EQ:EXPECTATION-BOUND}, we obtain
\begin{equation}
\label{EQ:BERNOULLI-BOUND}
    n\cdot c\left(R_{S}^{\Pi}, R^{\Pi} \right)  
    \leq  
    \mathrm{KL}\left(\Pi \| \pi_0\right)+\ln \left[\frac{1}{\epsilon}\cdot \mathrm{E}_{P^{n}} \left(\mathrm{E}_{\pi_{0}}\left(\exp\left[n\cdot c\left(R_{S}\left(g\right), R\left(g\right)\right)\right]\right)\right)\right].
\end{equation}
We now look to simplify $\mathrm{E}_{P^{n}} \left(\mathrm{E}_{\pi_{0}}\left( \exp\left[n\cdot c\left(R_{S}\left(g\right), R\left(g\right)\right)\right]\right)\right)$.
\begin{namedtheorem}[\textbf{\citet[\S Lemma 3]{Maurer2004}}]
Let $U$ denote a vector of $n$ independent and identically distributed random variables, each with values in the unit interval and $\mathrm{E}\left(U\right)=\mu$. 
In addition, let $U^\dagger$ be a Bernoulli random variable satisfying $\mathrm{E}\left(U^\dagger\right)=\mu$. 
Suppose that $b : \{0,1\}^{n}\rightarrow\mathbb{R}$ is a convex function. 
Then
\begin{equation}
    \mathrm{E}\left(b\left(U\right)\right)
    \leq 
    \mathrm{E}\left(b\left(U^\dagger\right)\right).
    \label{EQ:MAURER1}
\end{equation}
If $t$ is permutation-symmetric in its arguments and $\theta\left(k\right)$ denotes the $n$-dimensional binary vector whose first $k$ coordinates are one and whose remaining $n-k$ coordinates are zero, we also have
\begin{equation}
    \mathrm{E}\left(b\left(U^{\dagger}\right)\right) 
    = 
    {\sum_{k=0}^{n}}{n\choose k}\cdot\left(1-\mu\right)^{n-k}\cdot\mu^{k}\cdot b\left(\theta\left(k\right)\right).
    \label{EQ:MAURER2}
\end{equation}
\end{namedtheorem}
We then note that, by defining $b\left(U^\dagger\right) = \frac{1}{n}\sum_{i=1}^{n}U_{i}^\dagger$ such that $\mathrm{E}_{P^{n}}\left(b\left(U^\dagger\right)\right)=\mathrm{E}_{P^{n}}\left(R_{S}\left(g\right)\right)=R\left(g\right)$, since $c\left(\cdot,R\left(g\right)\right)$ is convex, and since the exponential function is convex and non-decreasing, $\exp\left[c\left(\cdot,R\left(g\right)\right)\right]$ is convex.
Applying \cref{EQ:MAURER1}, we obtain
\begin{equation}
\mathrm{E}_{P^{n}}\left( \mathrm{E}_{\pi_{0}} \left(\exp\left[n\cdot c\left(R_{S}\left(g\right), R\left(g\right)\right)\right]\right)\right)
\leq  
\mathrm{E}_{P^{n}}\left(\mathrm{E}_{\pi_{0}}\left(\exp\left[n\cdot c\left(b\left(U^{\dagger}\right), R\left(g\right)\right)\right]\right)\right),
\end{equation}
into which we substitute the definition of the expectation to write
\begin{equation}
\begin{aligned}
   \mathrm{E}_{P^{n}}\left(\mathrm{E}_{\pi_{0}} \left(\exp\left[n\cdot c\left(R_{S}\left(g\right), R\left(g\right)\right)\right]\right)\right)
   &\leq 
   \mathrm{E}_{\pi_{0}}\left(\sum_{k}P^n\left( b\left(U^{\dagger}\right) = \frac{k}{n}\right)\cdot \exp\left[n\cdot c\left(\frac{k}{n},R\left(g\right)\right)\right]\right),
   \end{aligned}
\end{equation}
where we omit the limits of summation to avoid confusion.
We then apply \cref{EQ:MAURER2} and the binomial theorem to obtain
\begin{align}
    &\mathrm{E}_{\pi_{0}}\left(\underset{k}{\overset{}{\sum}}P^n\left( b\left(U^\dagger\right) = \frac{k}{n}\right)\cdot \exp\left[n\cdot c\left(\frac{k}{n},R\left(g\right)\right)\right]\right)\\
    &= 
    \mathrm{E}_{\pi_{0}}\left(\underset{k=0}{\overset{n}{\sum}} {n \choose k}\cdot R\left(g\right)^{k}\cdot \left(1-R\left(g\right)\right)^{n-k}\cdot \exp\left[n\cdot c\left(\frac{k}{n},R\left(g\right)\right)\right]\right),\\
    &= 
    \mathrm{E}_{\pi_{0}}\left(\sum_{k=0}^{n} \mathrm{Bin}_{k}^{n}\left[R\left(g\right)\right)\cdot \exp\left(n\cdot c\left(\frac{k}{n},R\left(g\right)\right)\right]\right),\\
    &\leq 
    \sup _{r \in[0,1]}\,\sum_{k=0}^{n} \mathrm{Bin}_{k}^{n}\left(r\right)\cdot \exp\left({n\cdot c\left(\frac{k}{n}, r\right)}\right). 
\end{align}
If
\begin{equation}
\begin{aligned}
    c\left(q,p\right)
    &=
    \mathrm{KL}\left(q\| p\right) ,\\
    &=
    q\cdot \mathrm{ln}\left(\frac{q}{p}\right) + \left(1-q\right)\cdot\mathrm{ln}\left(\frac{1-q}{1-p}\right),
    \end{aligned}
\end{equation}
then
\begin{align}
    \sup _{r \in[0,1]}\sum_{k=0}^{n} \mathrm{Bin}_{k}^{n}\left(r\right)\cdot \exp\left({n\cdot c\left(\frac{k}{n}, r\right)}\right)
    &= 
    \sup _{r \in[0,1]}\sum_{k=0}^{n} {n \choose k}\cdot\left(\frac{k}{n}\right)^k\cdot\left(1-\frac{k}{n}\right)^{n-k},\\
    &= 
    \sum_{k=0}^{n} {n \choose k}\cdot\left(\frac{k}{n}\right)^k\cdot\left(1-\frac{k}{n}\right)^{n-k},\\
    &\equiv 
    \mathscr{I}_{KL},
\end{align}
which is a function of $n$ only.
\begin{namedtheorem}[\textbf{\citet[\S Theorem 1]{Maurer2004}}]
For all $n\geq 2$,
\begin{equation}
    \mathrm{E}\left( \exp\left[n\cdot \mathrm{KL}\left(R_{S}\left(g\right)\| R\left(g\right)\right)\right]\right)
    \leq 
    \exp\left(\frac{1}{12n}\right)\cdot\sqrt{\frac{\pi n}{2}}+2.
\end{equation}
\end{namedtheorem}
We emphasise that, here, $\pi$ signifies the mathematical constant rather than a probability distribution.
As is discussed in \citet{Maurer2004}, for all $n \geq 8$,
\begin{equation}
     \exp\left(\frac{1}{12n}\right)\cdot\sqrt{\frac{\pi n}{2}}+2
     \leq 
     2\sqrt{n}.
\end{equation}
Hence, for all $n \geq 8$ we have $\mathscr{I}_{KL}\leq2\sqrt{n}$. 
We are of course interested in the maximum distance between the expected welfare risk and its empirical analogue and so define $c_{V^{2}}\left(q,p\right) = 2\left(q-p\right)^{2}$.
Finally, we substitute $c_{V^{2}}\left(q,p\right)$ into \cref{EQ:BERNOULLI-BOUND} and utilise Pinsker's Inequality, which states that $c_{V^{2}}\left(q,p\right)\leq \mathrm{KL}\left(q\| p\right)$, together with \citet[\S Theorem 1]{Maurer2004} to obtain
\begin{align}
    R^{\Pi}-R^{\Pi}_{S}
    &\leq
    \sqrt{\frac{1}{2n}\cdot\mathrm{KL}\left(\Pi \| \pi_0\right)+\frac{1}{2n}\cdot\ln \left[\frac{1}{\epsilon}\cdot \mathrm{E}_{P^{n}}\left(\mathrm{E}_{\pi_{0}}\left( \exp\left[n\cdot c_{V^{2}}\left(R_{S}\left(g\right), R\left(g\right)\right)\right]\right)\right)\right]},\\
    &\leq
    \sqrt{\frac{1}{2n}\cdot\mathrm{KL}\left(\Pi \| \pi_0\right)+\frac{1}{2n}\cdot\ln \left[\frac{1}{\epsilon}\cdot \mathrm{E}_{P^{n}}\left( \mathrm{E}_{\pi_{0}}\left(\exp\left[n\cdot \mathrm{KL}\left(R_{S}\left(g\right)\| R\left(g\right)\right)\right]\right)\right)\right]},\\
    &\leq
    \sqrt{\frac{1}{2n}\cdot\mathrm{KL}\left(\Pi \| \pi_0\right)+\frac{1}{2n}\cdot\ln\left(\frac{1}{\epsilon}\cdot\mathscr{I}_{KL}\right)},\\
    &\leq
    \sqrt{\frac{1}{2n}\left(\mathrm{KL}\left(\Pi \| \pi_0\right)+ \ln\left(\frac{2\sqrt{n}}{\epsilon}\right)\right)}.
\end{align}
The desired result follows by rearrangement.


\subsection{Proof of \cref{THM:OPTIMAL}}
\label{PROOF:OPTIMAL}

An optimal posterior minimises
\begin{equation}
\label{EQ:CONSTRAINED-PROBLEM}
\begin{aligned}
\mathrm{E}_{\Pi}\left(R_{S}\left(g\right)\right) + \sqrt{\frac{1}{2n}\cdot\left( \mathrm{E}_{\Pi}\left(\ln\left(\frac{\mathrm{d}\Pi\left(g\right)}{{\mathrm{d}\pi_{0}}\left(g\right)}\right)\right) + \ln\left(\frac{2\sqrt{n}}{\epsilon}\right)  \right)}\text{ subject to } \int_{\mathscr{G}} \mathrm{d}\Pi\left(g\right)=1,&\\
\inf_{g\in\mathscr{G}}\,\mathrm{d}\Pi\left(g\right)\geq 0.&
\end{aligned}
\end{equation}
Provided that the solution satisfies the non-negativity constraints (the second constraint of \cref{EQ:CONSTRAINED-PROBLEM}), this is equivalent to minimising
\begin{equation} 
\label{EQ:LAGRANGE-1}
\int_{\mathscr{G}} R_{S}\left(g\right)\cdot\mathrm{d}\Pi\left(g\right) + \sqrt{\frac{1}{2n}\cdot\left( \int_{\mathscr{G}} \ln\left(\frac{\mathrm{d}\Pi\left(g\right)}{{\mathrm{d}\pi_{0}}\left(g\right)}\right)\cdot\mathrm{d}\Pi\left(g\right) + \ln\left(\frac{2\sqrt{n}}{\epsilon}\right)  \right)}+\xi\cdot\left(\int_{\mathscr{G}}\mathrm{d}\Pi\left(g\right)-1\right),
\end{equation}
where $\xi$ is the Lagrange multiplier.
We separate this minimisation into two parts, by minimising \cref{EQ:LAGRANGE-1} over $\Pi$ subject to its Kullbuck-Leibler divergence from $\pi_0$ being equal to $c \geq 0$ and, subsequently, by searching for the $c$ that minimises the objective function. 

We can write the constrained minimisation that forms the first part of the problem as
\begin{equation} 
\label{EQ:LAGRANGE-2}
\int_{\mathscr{G}} R_{S}\left(g\right)\cdot\mathrm{d}\Pi\left(g\right) + \sqrt{\frac{1}{2n}\cdot\left(c + \ln\left(\frac{2\sqrt{n}}{\epsilon}\right)  \right)}+\xi \cdot \left(\int_{\mathscr{G}}\mathrm{d}\Pi\left(g\right)-1\right)+\frac{1}{\chi}\cdot\left( \int_{\mathscr{G}} \ln\left(\frac{\mathrm{d}\Pi\left(g\right)}{{\mathrm{d}\pi_{0}}\left(g\right)}\right)\cdot\mathrm{d}\Pi\left(g\right)-c\right),
\end{equation}
where $\xi$ and $1/\chi$ are Lagrange multipliers, and where $c\geq0$ is a constant, provided that the omitted non-negativity constraints are satisfied at the solution.
The associated first order condition of the minimand with respect to $\mathrm{d}\Pi\left(g\right)$ is
\begin{equation}
R_{S}\left(g\right) +\xi+\frac{1}{\chi}\cdot\left(\ln\left(\frac{\mathrm{d}\Pi\left(g\right)}{\mathrm{d}\pi_{0}\left(g\right)}\right)+1\right)
=
0.
\end{equation}
Rearranging, we obtain
\begin{equation}
\mathrm{d}\Pi^{*}\left(g\right) 
= 
\frac{\exp\left({-\chi \cdot R_{S}\left(g\right) }\right)}{ \exp\left({1+\xi\cdot\chi}\right)}\cdot\mathrm{d}{\pi_{0}}\left(g\right),
\end{equation}
where we emphasise that $\Pi^{*}$ is a function of $\chi$.
In view of the first constraint of \cref{EQ:LAGRANGE-1},
\begin{equation}
    \exp\left({1+\xi\cdot\chi}\right)
    =
    \int_{\mathscr{G}}\exp\left({-\chi \cdot R_{S}\left(g\right) }\right)\cdot\mathrm{d}\pi_{0}\left(g\right),
\end{equation}
and so, for all $g\in\mathscr{G}$,
\begin{equation}
\label{EQ:LAGRANGE-3}
    \mathrm{d}\Pi^{*}\left(g\right)
=
\frac{\exp \left(-\chi\cdot R_{S}\left(g\right)\right)}{\int_{\mathscr{G}}\exp\left(-\chi\cdot R_{S}\left(g\right)\right)\cdot\mathrm{d}{\pi_{0}}\left(g\right)}\cdot{\mathrm{d}\pi_{0}\left(g\right)},
\end{equation}
which integrates to one as is required.
We reiterate that \cref{EQ:LAGRANGE-3} is derived for an arbitrary $c\geq0$, and so holds for any feasible values of $\chi$.

For $\Pi^{*}$ to satisfy $\mathrm{KL}\left(\Pi^{*} \| \pi_0 \right) = c$, we require that
\begin{equation}
\begin{aligned}
    c
    &= 
    \int_{\mathscr{G}}\ln\left(\frac{\mathrm{d}\Pi^{*}\left(g\right)}{\mathrm{d}\pi_{0}\left(g\right)}\right)\cdot\mathrm{d}\Pi^{*}\left(g\right),\\
    &=
    -\int_{\mathscr{G}}\chi\cdot R_{S}\left(g\right)\cdot\mathrm{d}\Pi^{*}\left(g\right)-\ln\left(\int_{\mathscr{G}}\exp\left(-\chi\cdot R_{S}\left(g\right)\right)\cdot\mathrm{d}{\pi_{0}}\left(g\right)\right).\\
   \end{aligned} 
\end{equation}
This relationship between the radius of the Kullback-Leibler ball and the inverse of the Lagrange multiplier shows that $c$ is strictly monotonically increasing in $\chi > 0$ provided that $R_{S}\left(g\right)$ is not constant over those $g$ supported by $\pi_0$, since
\begin{align}
    \frac{\mathrm{d}c}{\mathrm{d}\chi} & = -\int_{\mathscr{G}} R_{S}\left(g\right)\cdot\mathrm{d}\Pi^{*}\left(g\right) - \int_{\mathscr{G}}\chi\cdot R_{S}\left(g\right)\cdot \frac{\mathrm{d}}{\mathrm{d}\chi} \left( \frac{\mathrm{d}\Pi^{*}\left(g\right)}{\mathrm{d}\pi_{0}\left(g\right)}\right)\cdot\mathrm{d}\pi_{0} +\int_{\mathscr{G}} R_{S}\left(g\right)\cdot\mathrm{d}\Pi^{*}\left(g\right) \\
    &= 
    \chi\cdot\int_{\mathscr{G}}\left(R_{S}\left(g\right)-R_{S}^{\Pi^{*}}\right)^{2}\cdot\mathrm{d}\Pi^{*}\left(g\right)\\
    &\geq 
    0,
    \label{EQ:STRICT}
\end{align}
where \cref{EQ:STRICT} is strict if $\chi > 0$ and $R_{S}\left(g\right)\neq R_{S}^{\Pi^{*}}$ for at least some $g$ supported by $\pi_{0}$. 
Accordingly, in the second part of the optimisation, we substitute $\Pi^{*}$ into \cref{EQ:LAGRANGE-2} in place of $\Pi$ and solve  the unconstrained minimisation problem with respect to $\chi$.
That is, we solve
\begin{equation}
    \label{EQ:LAGRANGE-4}
\min_{\chi} \left\{ 
R_{S}^{\Pi^{*}} + \underbrace{\sqrt{\frac{1}{2n}\cdot\left(-\chi\cdot R_{S}^{\Pi^{*}}-\ln\left(\int_{\mathscr{G}}\exp\left(-\chi\cdot R_{S}\left(g\right)\right)\cdot\mathrm{d}\pi_{0}\left(g\right)\right) + \ln\left(\frac{2\sqrt{n}}{\epsilon}\right)  \right)}}_{\textrm{Penalty}} 
\right\}.
\end{equation}
We note that
\begin{equation}
\begin{aligned}
    \frac{\mathrm{d}}{\mathrm{d}\chi} R^{\Pi^{*}}_{S}
    &=
    \frac{\mathrm{d}}{\mathrm{d}\chi}\int_{\mathscr{G}}R_{S}\left(g\right)\cdot\frac{\exp \left(-\chi\cdot R_{S}\left(g\right)\right)}{\int_{\mathscr{G}}\exp\left(-\chi\cdot R_{S}\left(g\right)\right)\cdot\mathrm{d}{\pi_{0}}\left(g\right)}\cdot{\mathrm{d}\pi_{0}\left(g\right)},\\
    &=
    \int_{\mathscr{G}}R_{S}\left(g\right)\cdot\frac{\mathrm{d}}{\mathrm{d}\chi}\frac{\exp \left(-\chi\cdot R_{S}\left(g\right)\right)}{\int_{\mathscr{G}}\exp\left(-\chi\cdot R_{S}\left(g\right)\right)\cdot\mathrm{d}{\pi_{0}}\left(g\right)}\cdot{\mathrm{d}\pi_{0}\left(g\right)},\\
    &=
    -\int_{\mathscr{G}}\left(R_{S}\left(g\right)-R_{S}^{\Pi^{*}}\right)^{2}\cdot\mathrm{d}\Pi^{*}\left(g\right),
    \end{aligned}
    \label{EQ:FOC-1}
\end{equation}
which we interpret as the variance of empirical welfare risk under $\Pi^{*}$.
Using this result, we further note that
\begin{equation}
    \begin{aligned}
\frac{\mathrm{d}}{\mathrm{d}\chi}\textrm{Penalty}
&=
\frac{1}{4n\cdot\textrm{Penalty}}\cdot\left(-R_{S}^{\Pi^{*}}-\chi\cdot\frac{\mathrm{d}}{\mathrm{d}\chi}R^{\Pi^{*}}_{S}+\frac{\int_{\mathscr{G}}R_{S}\left(g\right)\cdot\exp\left(-\chi\cdot R_{S}\left(g\right)\right)\cdot\mathrm{d}\pi_{0}\left(g\right)}{\int_{\mathscr{G}}\exp\left(-\chi\cdot R_{S}\left(g\right)\right)\cdot\mathrm{d}\pi_{0}\left(g\right)}\right),\\
&=
\frac{1}{4n\cdot\textrm{Penalty}}\cdot-\chi\cdot\frac{\mathrm{d}}{\mathrm{d}\chi}R^{\Pi^{*}}_{S},\\
&=
\frac{1}{4n\cdot\textrm{Penalty}}\cdot\chi\cdot\int_{\mathscr{G}}\left(R_{S}\left(g\right)-R_{S}^{\Pi^{*}}\right)^{2}\cdot\mathrm{d}\Pi^{*}\left(g\right).
\end{aligned}
\label{EQ:FOC-2}
\end{equation}
Together \cref{EQ:FOC-1,EQ:FOC-2} imply that the associated first order condition of \cref{EQ:LAGRANGE-4} with respect to $\chi$ is
\begin{equation}
    \frac{1}{4n\cdot\textrm{Penalty}}\cdot\chi\cdot\int_{\mathscr{G}}\left(R_{S}\left(g\right)-R_{S}^{\Pi^{*}}\right)^{2}\cdot\mathrm{d}\Pi^{*}\left(g\right)
    =
    \int_{\mathscr{G}}\left(R_{S}\left(g\right)-R_{S}^{\Pi^{*}}\right)^{2}\cdot\mathrm{d}\Pi^{*}\left(g\right).
\end{equation}
Provided -- trivially, we might add -- that $\Pi^{*}$ is not degenerate and there is variation in empirical welfare risk (the conditions under which \cref{EQ:STRICT} is strict), then this condition reduces to 
\begin{equation}
\label{EQ:FIXED-POINT}
\chi 
=
4 n\cdot\sqrt{\frac{1}{2 n}\cdot\left(\mathrm{KL}\left(\Pi^{*} \| \pi_0\right)+\ln\left( \frac{2 \sqrt{n}}{\epsilon}\right)\right)} ,
\end{equation}
which is exactly the condition that appears in \cref{THM:OPTIMAL}.

Although \cref{EQ:FIXED-POINT,EQ:LAGRANGE-3} characterise a possible interior solution of the optimisation, we have yet to guarantee that this proposed solution is a global optimum.
To address this, we show that the two possible corner solutions are sub-optimal, such that the first order conditions from which \cref{EQ:FIXED-POINT} are derived are applicable. 
Continuity and differentiability of \cref{EQ:LAGRANGE-2} are then sufficient\footnote{%
It is possible to show that the difference between the left- and right-hand sides of \cref{EQ:FIXED-POINT} is negative at $\chi=0$ and positive if $\chi$ is sufficiently large, with existence then established using extensions of the intermediate value theorem and its corollary, Bolzano's theorem.} 
to guarantee that a fixed point satisfying \cref{EQ:FIXED-POINT} exists.

Neither $\pi_{0}$ nor a degenerate distribution are optimal.
To establish that $\pi_{0}$ is not optimal, we note that $\Pi^{*}$ coincides with $\pi_{0}$ when $\chi=0$.
By marginally increasing $\chi$, such that we move in the direction of $\Pi^{*}$, we obtain a probability distribution that is in the interior.
It suffices to show that such a probability distribution reduces the value of the objective function relative to $\pi_{0}$.
Evaluating \cref{EQ:FOC-1,EQ:FOC-2} at $\chi=0$, we obtain
\begin{equation}
\begin{aligned}
    \frac{\mathrm{d}}{\mathrm{d}\chi}\left.\left(R^{\Pi^{*}}_{S}+\textrm{Penalty}\right)\right\vert_{\chi=0}
    &=
    \frac{\mathrm{d}}{\mathrm{d}\chi}R^{\pi_{0}}_{S},\\
    &<0,
    \end{aligned}
\end{equation}
as we require, with any $\chi$ satisfying $0<\chi<2n\cdot R^{\pi_{0}}$ yielding a strictly lower value than $\chi=0$.
To establish that a degenerate distribution is not optimal, we note that such a distribution implies infinite divergence from $\pi_{0}$ (if $\pi_{0}$ were itself degenerate then our analysis would be meaningless since the prior and posterior distributions would always coincide; and if $\pi_{0}$ is atomic then the discrete case would apply).
Given that $\pi_{0}$ attains a finite value of the objective function, however, $\pi_{0}$ is always preferred to a degenerate distribution, and so a degenerate distribution cannot be optimal.


\subsection{Proof of \cref{THM:VARIATIONAL}}
\label{PROOF:VARIATIONAL}

In the course of this proof, we rely on several results in \citet{Alquier2016} and follow the standard PAC-Bayesian approach (see \citealp{catoni2007pac}).
We define
\begin{equation}
    \mathrm{d}\Pi^{*}\left(g\right)
=
\frac{\exp \left[-\lambda\cdot \left(R\left(g\right)-R_{S}\left(g\right)\right)\right]}{\int_{\mathscr{G}}\exp\left[-\lambda\cdot \left(R\left(g\right)-R_{S}\left(g\right)\right)\right]\cdot\mathrm{d}{\pi_{0}}\left(g\right)}\cdot{\mathrm{d}\pi_{0}\left(g\right)},
\end{equation}
which differs from \cref{EQ:LAGRANGE-3} in its use of the notation $\lambda$ rather than $\chi$ and in terms of the measures of risk that it uses, such that
\begin{equation}\label{EQ:LOG-EQUATION}
    \ln\left(\int_{\mathscr{G}} \exp \left[-\lambda\cdot  \left(R\left(g\right)-R_{S}\left(g\right)\right)\right]\cdot \mathrm{d} {\pi_{0}\left(g\right)}\right)
    =
    -\inf_{\Pi \in \mathscr{M}}\,\left\{\int_{\mathscr{G}} \lambda\cdot\left(R\left(g\right)-R_{S}\left(g\right)\right)\cdot \mathrm{d} \Pi\left(g\right)+\mathrm{KL}\left(\Pi \| {\pi_{0}}\right)\right\},
\end{equation}
by definition of the Kullback-Leibler divergence and of $\Pi^{*}$.

As in \citet{Alquier2016}, we define a Hoeffding assumption as
\begin{equation}
\label{EQ:HOEFFDING-1}
    \mathrm{E}_{\pi_{0}}\left(\mathrm{E}_{P^{n}}\left( \exp\left[\lambda\cdot\left(R\left(g\right)-R_{S}\left(g\right)\right)\right]\right)\right)
    \leq 
    \exp\left(a\left(\lambda, n\right)\right),
\end{equation}
where $a$ is a function that depends upon $\lambda>0$. This follows directly from applying Hoeffding's lemma \citep{boucheron2013concentration} to the left-hand side of \cref{EQ:HOEFFDING-1} and then taking the expectation over $\mathscr{G}$ with respect to $\pi_{0}$.  
As we have bounded the welfare criterion such that it is contained in the unit interval, this assumption holds directly without any further restrictions.
Applying Fubini's Theorem to \cref{EQ:HOEFFDING-1}, we obtain
\begin{equation}
\label{EQ:HOEFFDING-2}
    \mathrm{E}_{P^{n}}\left( \int  \exp \left(\lambda\cdot \left(R_{S}\left(g\right)-R\left(g\right)\right) - a\left(\lambda,n\right)\right)\cdot d{\pi_{0}}\right)
    \leq
    1.
\end{equation}
Given that (\ref{EQ:LOG-EQUATION}) can be re-stated as
\begin{equation}
\label{EQ:LOG-EQUATION2}
    \int_{\mathscr{G}} \exp \left[-\lambda\cdot \left(R\left(g\right)-R_{S}\left(g\right)\right)\right]\cdot \mathrm{d} {\pi_{0}\left(g\right)}
    =
    \exp\left(\sup _{\Pi \in \mathscr{M}}\,\int_{\mathscr{G}}\lambda\cdot\left(R_{S}\left(g\right)-R\left(g\right)\right)\cdot \mathrm{d} \Pi\left(g\right)-\mathrm{KL}\left(\Pi \| {\pi_{0}}\right)\right),
\end{equation}
it follows from \cref{EQ:LOG-EQUATION,EQ:HOEFFDING-2} that 
\begin{equation}
\label{proofeq:lem1inequality1}
    \mathrm{E}_{P^{n}}\left[\exp\left(\sup_{\Pi\in\mathscr{M}}\,\lambda\cdot\int_{\mathscr{G}}\left(R_{S}\left(g\right)-R\left(g\right)\right)\cdot\mathrm{d}\Pi\left(g\right)-\mathrm{KL}\left(\Pi\|\pi_{0}\right)-a\left(\lambda,n\right)\right)\right]
    \leq
    1.
\end{equation}
Multiplying both sides of \cref{proofeq:lem1inequality1} by $\epsilon/2$ and using the fact that $\mathrm{E}\left(\exp\left(U\right)\right)\geq\Pr\left(U>0\right)$ for any random variable $U$, we obtain
\begin{equation}
\label{EQ:BONFERRONI-1}
    P^n\left(\sup_{\Pi\in\mathscr{M}}\, \lambda\cdot\int_{\mathscr{G}} \left(R_{S}\left(g\right) - R\left(g\right)\right) \cdot\mathrm{d}\Pi\left(g\right) -  \mathrm{KL}\left(\Pi||{\pi_{0}}\right) -a\left(\lambda,n\right) + \ln\left(\frac{\epsilon}{2}\right)   > 0\right)
    \leq 
    \frac{\epsilon}{2}.
\end{equation}
Keeping this bound in mind, we now turn to the problem of bounding expected welfare regret.

We let $c$ denote a convex function that maps from the unit square to the reals. 
Then, by Jensen's Inequality,
\begin{equation}
\label{EQ:JENSEN}
\begin{aligned}
    n\cdot c\left(R^{\Pi}_{S},R^{\Pi}\right) 
    &=
    n\cdot c\left[\mathrm{E}_{\Pi}\left( R_{S}\left(g\right)\right),\mathrm{E}_{\Pi}\left( R\left(g\right)\right)\right],\\
    &\leq 
    \mathrm{E}_{\Pi} \left[n\cdot c\left(R_{S}\left(g\right),R\left(g\right)\right)\right].
    \end{aligned}
\end{equation}
Letting $\phi\left(g\right) =  n\cdot c\left(R_{S}\left(g\right),R\left(g\right)\right)$, we apply \citet[\S Lemma 3]{Begin2016} to obtain
\begin{equation}
     \mathrm{E}_{\Pi}\left[ n\cdot c\left(R_{S}\left(g\right),R\left(g\right)\right)\right] 
     \leq 
     \mathrm{KL}\left(\Pi \| {\pi_{0}}\right)+\ln\left(\mathrm{E}_{\pi_{0}}\left(\exp\left[n\cdot c\left(R_{S}\left(g\right),R\left(g\right)\right)\right]\right)\right),
\end{equation}
which, in conjunction with \cref{EQ:JENSEN}, yields
\begin{equation}
\label{EQ:APPLY-BEGIN}
   n\cdot c\left(R^{\Pi}_{S},R^{\Pi}\right)
   \leq 
   \mathrm{KL}\left(\Pi \| {\pi_{0}}\right)+\ln\left(\mathrm{E}_{\pi_{0}}\left(\exp\left[n\cdot c\left(R_{S}\left(g\right),R\left(g\right)\right)\right]\right)\right).
\end{equation}
We then apply Markov's inequality to the term inside the logarithm, which yields
\begin{equation}
\label{EQ:BONFERRONI-2}
    P^n\left[\mathrm{E}_{\pi_{0}}\left(\exp\left(n\cdot c\left[R_{S}\left(g\right),R\left(g\right)\right)\right]\right) \geq \frac{2}{\epsilon}\cdot \mathrm{E}_{P^{n}}\left(\mathrm{E}_{\pi_{0}}\left( \exp\left[ n\cdot c\left(R_{S}\left(g\right),R\left(g\right)\right)\right]\right)\right)\right]
    \leq 
    \frac{\epsilon}{2}.
\end{equation}
Combining \cref{EQ:BONFERRONI-1,EQ:BONFERRONI-2} by means of Bonferroni's inequalities, we determine that, for all $\Pi \in \mathscr{M}$,
\begin{equation}
\label{EQ:UB-1}
    \mathrm{E}_{\pi_{0}}\left(\exp\left[n\cdot c\left(R_{S}\left(g\right),R\left(g\right)\right)\right]\right)
    \leq \frac{2}{\epsilon}\cdot \mathrm{E}_{P^{n}}\left(\mathrm{E}_{\pi_{0}}\left(\exp\left[n\cdot c\left(R_{S}\left(g\right),R\left(g\right)\right)\right]\right)\right),
\end{equation}
and
\begin{equation}
     \lambda\cdot\mathrm{E}_{\Pi}\left(R_{S}\left(g\right)\right) 
     \leq 
     \lambda\cdot\mathrm{E}_{\Pi}\left(R\left(g\right)\right) + \mathrm{KL}\left(\Pi||{\pi_{0}}\right) + a\left(\lambda,n\right) + \ln\left(\frac{2}{\epsilon}\right),
\end{equation}
which both hold simultaneously with probability at least $1-\epsilon$.
We then take the exponential of both sides of \cref{EQ:APPLY-BEGIN} and rearrange, relating the result to the left-hand side of \cref{EQ:UB-1} to obtain
\begin{equation}
\label{EQ:PRELIMINARY} 
n\cdot c\left(R_{S}^{\Pi}, R^\Pi\right)  
\leq  
\mathrm{KL}\left(\Pi \| {\pi_{0}}\right)+\ln \left[\frac{2}{\epsilon}\cdot \mathrm{E}_{P^{n}}\left(\mathrm{E}_{\pi_{0}}\left(\exp\left[n\cdot c\left(R_{S}\left(g\right), R\left(g\right)\right)\right]\right)\right)\right].
\end{equation}
We note that \cref{EQ:PRELIMINARY} is identical to \cref{EQ:BERNOULLI-BOUND} except for the logarithm on the right hand-side, the argument of which differs only in the constant that multiplies the expectation.
As such, and following the steps of \cref{PROOF:PAC} that follow \cref{EQ:BERNOULLI-BOUND}, we are able to show that, for all $\Pi\in\mathscr{M}$,
\begin{equation}
\label{EQ:APPLY-LEMMA-1}
    R^{\Pi} 
    \leq 
    R_{S}^{\Pi} + \sqrt{\frac{1}{2n}\cdot\left( \mathrm{KL}\left(\Pi||{\pi_{0}}\right) + \ln \left( \frac{4\sqrt{n}}{\epsilon} \right) \right) }, 
\end{equation}
and
\begin{equation} 
\label{EQ:APPLY-LEMMA-2}
    \lambda R_{S}^{\Pi} 
    \leq 
    \lambda\cdot  R^{\Pi} +  \mathrm{KL}\left(\Pi||{\pi_{0}}\right) +a\left(\lambda,n\right) + \ln\left(\frac{2}{\epsilon}\right),
\end{equation}
which both hold simultaneously with probability at least $1-\epsilon$.
We now denote by $\mathscr{V}\subset\mathscr{M}$ a family of distributions over the class of feasible assignment rules and restrict attention to this subset.
Since $\mathscr{V}\subset\mathscr{M}$, \cref{EQ:APPLY-LEMMA-1,EQ:APPLY-LEMMA-2} hold with probability at least $1-\epsilon$ for all $\Pi\in\mathscr{V}$ and, in particular, for the variational approximation of $\Pi^*$, which we define as 
\begin{equation}
    \tilde{\Pi}
    =
    \argmin_{\Pi\in\mathscr{G}}\,\mathrm{KL}\left(\Pi\|\Pi^*\right).
\end{equation}
Hence, from \cref{EQ:APPLY-LEMMA-1}, 
\begin{equation} 
\label{EQ:APPLY-LEMMA-3}
    R^{\tilde{\Pi}} 
    \leq 
    \inf_{\Pi\in\mathscr{V}}\,R_{S}^{\Pi} + \sqrt{\frac{1}{2n}\left( \mathrm{KL}\left(\Pi||{\pi_{0}}\right) + \ln \left( \frac{4\sqrt{n}}{\epsilon} \right) \right)}.
\end{equation}
Finally, dividing both sides of \cref{EQ:APPLY-LEMMA-2} by $\lambda$ and replacing $R^{\Pi}_{S}$ inside the infimum of \cref{EQ:APPLY-LEMMA-3} with the right-hand side of \cref{EQ:APPLY-LEMMA-2}, we obtain the required result, which we reiterate holds with probability at least $1-\epsilon$.


\subsection{Proof of \cref{THM:RATE}}
\label{PROOF:RATE}

We rely on an intermediate result that we introduce now, in advance of the proof.
\begin{lemma}
\label{THM:CIRCULAR}
The circular variance of a von Mises-Fisher random vector with concentration parameter $\kappa\geq 0$ is bounded from above by
\begin{equation}
    \frac{\underline{c}_{\nu}+\sqrt{\kappa^{2}+\overline{c}_{\nu}^{2}}-\kappa}{\underline{c}_{\nu}+\sqrt{\kappa^{2}+\overline{c}_{\nu}^{2}}},
\end{equation}
where $\nu=m/2-1$ for convenience, with $\underline{c}_{\nu}=\nu+1/2$ and $\overline{c}_{\nu}=\nu+3/2$.
\end{lemma}
As is shown in \citet{KitagawaRowley}, the circular variance is an $\mathrm{O}\left(\kappa^{-1}\right)$ function.
Proof of \cref{THM:CIRCULAR} is deferred until the end of this proof.
\begin{lemma}
\label{THM:KL-RATE}
The Kullback-Leibler divergence of a von Mises-Fisher random vector with concentration parameter $\kappa\geq 0$ is bounded from above by
\begin{equation}
\label{EQ:KL-BOUND}
    \underline{c}_{\nu}\cdot\ln\left(\frac{\underline{c}_{\nu}+\sqrt{\kappa^{2}+\overline{c}_{\nu}^{2}}}{\underline{c}_{\nu}+\overline{c}_{\nu}}\right)+\sqrt{\vphantom{\overline{c}_{\nu}^{2}}\kappa^{2}+\underline{c}_{\nu}^{2}}-\sqrt{\kappa^{2}+\overline{c}_{\nu}^{2}}+1,
\end{equation}
where $\nu=m/2-1$ for convenience, with $\underline{c}_{\nu}=\nu+1/2$ and $\overline{c}_{\nu}=\nu+3/2$.
\end{lemma}
As is shown in \citet{KitagawaRowley}, the Kullback-Leibler divergence is an $\mathrm{O}\left(\ln\left(\kappa\right)\right)$ function.
Proof of \cref{THM:KL-RATE} is deferred until the end of this proof.

With \cref{THM:CIRCULAR,THM:KL-RATE} to hand, we begin by noting that, together, \cref{ASS:CO,THM:VARIATIONAL} imply that
\begin{equation}
\label{EQ:APPLY-BEGIN'S-THEOREM}
        R^{\tilde{\Pi}} 
        \leq 
        \bar{R} + \inf_{\mu,\kappa}a_{1}+a_{2}+\sqrt{a_{3}},
\end{equation}
where 
\begin{equation}
\begin{aligned}
    a_{1}
    &=
    \int_{\mathbb{S}^{m-1}} 2c\cdot\|\beta-\overline{\beta}\|_{2}\cdot\mathrm{d}\Pi\left(\beta;\kappa,\mu\right)\vphantom{\left(\frac{1}{\lambda}\right)},\\
    a_{2}
    &=
    \frac{1}{\lambda}\cdot\left(a\left(\lambda,n\right)+\mathrm{KL}\left(\Pi||{\pi_{0}}\right)+\ln\left(\frac{2}{\epsilon}\right)\right),\\
    a_{3}
    &=
    \frac{1}{2n}\cdot\left( \mathrm{KL}\left(\Pi||{\pi_{0}}\right) + \ln \left( \frac{4\sqrt{n}}{\epsilon} \right) \right),
    \end{aligned}
\end{equation}
and where $\mathbb{S}^{m}$ is the $m$-sphere (i.e., the hypersphere in $\mathbb{R}^{m+1}$).

Focusing on $a_{1}$, we use Jensen's inequality to show that
\begin{align}
    2c\cdot \int_{\mathbb{S}^{m-1}} \|\beta - \overline{\beta} \|_{2}\cdot\mathrm{d}\Pi\left(\beta;\kappa,\mu\right) 
    &\leq 
    2c \cdot\sqrt{\int_{\mathbb{S}^{m-1}} \sum_{l=0}^{m}\left(\beta_{l}-\overline{\beta}_{l}\right)^{2}\cdot\mathrm{d}\Pi\left(\beta;\kappa,\mu\right)} \\
    &= 
    2c\cdot\sqrt{\mathrm{tr}\left(\mathrm{E}_{\Pi}\left(\left(\beta-\overline{\beta}\right)\cdot\left(\beta-\overline{\beta}\right)^{\top}\right)\right)}.
\end{align}
Rearranging, we obtain
\begin{align}
    \mathrm{E}_{\Pi}\left(\left(\beta-\overline{\beta}\right)\cdot\left(\beta-\overline{\beta}\right)^{\top}\right) 
    &= 
    \mathrm{E}_{\Pi}\left(\left(\beta-\overline{\beta}+\mathrm{E}_{\Pi}\left(\beta\right)-\mathrm{E}_{\Pi}\left(\beta\right)\right)\cdot\left(\beta-\overline{\beta}+\mathrm{E}_{\Pi}\left(\beta\right)-\mathrm{E}_{\Pi}\left(\beta\right)\right)^{\top}\right)\\
    &= 
    \mathrm{E}_{\Pi}\left(\left(\beta-\mathrm{E}_{\Pi}\left({\beta}\right)\right)\cdot\left(\beta-\mathrm{E}_{\Pi}\left(\beta\right)\right)^{\top}\right) + \left(\mathrm{E}_{\Pi}\left(\beta\right)-\overline{\beta}\right)\cdot\left(\mathrm{E}_{\Pi}\left(\beta\right)-\overline{\beta}\right)^{\top}\\
    &= 
    \mathrm{Variance}_{\Pi}\left(\beta\right) + \left(A_{m}\left(\kappa\right)\cdot\mu - \overline{\beta}\right)\cdot\left(A_{m}\left(\kappa\right)\cdot\mu - \overline{\beta}\right)^{\top},
\end{align}
where 
\begin{equation}
    A_{m}\left(\kappa\right)
    =
    \frac{I_{m/2}\left(\kappa\right)}{I_{m/2-1}\left(\kappa\right)},
\end{equation}
which relies on two results from \citet{mardia2009directional} and \citet{KitagawaRowley}. 
These results relate to the first two moments of the von Mises-Fisher distribution and are
\begin{gather}
\mathrm{E}_{\Pi}\left(\beta\right)
=
A_{m}\left(\kappa\right)\cdot \mu,\\
    \mathrm{Variance}_{\Pi}\left(\beta\right)
    =
    \frac{1}{\kappa}\cdot A_{m}\left(\kappa\right)\cdot \mathbb{I}_{m}+\left(1-\frac{m}{\kappa}\cdot A_{m}\left(\kappa\right)-A_{m}^{2}\left(\kappa\right)\right)\cdot \mu\cdot \mu^{\prime},
\end{gather}
where $\mathbb{I}_{m}$ is the m-dimensional identity matrix.
Restricting the set of von Mises-Fisher distributions to those satisfying $\mu=\overline{\beta}$, $a_{1}$ is bounded by
\begin{equation}
\begin{aligned}
    2c\cdot\sqrt{\mathrm{tr}\left(\mathrm{E}\left(\left(\beta-\overline{\beta}\right)\cdot\left(\beta-\overline{\beta}\right)^{\top}\right)\right)}
    &=
    2c\cdot\sqrt{\mathrm{tr}\left(\mathrm{Variance}\left(\beta\right)+\left(A_{m}\left(\kappa\right)-1\right)^{2}\cdot\mu\cdot\mu^{\top} \right)} \\
    &= 
    \sqrt{8c^{2}}\cdot\sqrt{1-A_{m}\left(\kappa\right)},
    \end{aligned}
    \label{EQ:A1}
\end{equation}
which is proportional to the square root of the circular variance.\footnote{%
Recall that the right-hand side of \cref{EQ:APPLY-BEGIN'S-THEOREM} is preceded by an infimum over $\mu$; it is not necessary for $\mu=\overline{\beta}$ to minimise \cref{EQ:A1} for this choice to deliver a useful upper bound.}
\cref{THM:CIRCULAR} details the behaviour of the circular variance.
Using \cref{THM:CIRCULAR}, \cref{EQ:A1} can be bounded from above by
\begin{equation}
\label{EQ:SUBA1}
    \sqrt{8c^{2}\cdot\frac{\underline{c}_{\nu}+\sqrt{\kappa^{2}+\overline{c}_{\nu}^{2}}-\kappa}{\underline{c}_{\nu}+\sqrt{\kappa^{2}+\overline{c}_{\nu}^{2}}}}
    \leq
    4c\cdot\sqrt{\frac{\nu+1}{\kappa}},
\end{equation}
where the right-hand side follows from the definitions of $\underline{c}_{\nu}$ and $\overline{c}_{\nu}$ and concavity of the square root, and by decreasing the denominator via elimination of $\underline{c}_{\nu}$ and $\overline{c}_{\nu}$ (both are non-negative constants).
We substitute the right-hand side of \cref{EQ:SUBA1} into \cref{EQ:APPLY-BEGIN'S-THEOREM} in place of $a_{1}$.

We now focus on the Kullback-Leibler divergence, which appears in both $a_2$ and $a_3.$
We observe that the difference between the square roots in \cref{EQ:KL-BOUND} is increasing in $\kappa$ (i.e, the difference becomes less negative as $\kappa$ increases and as $\kappa$ increases in importance relative to $\nu$).
As such, it suffices to omit the difference between the square roots in \cref{EQ:KL-BOUND} and to simply bound the Kullback-Leibler divergence from above by
\begin{equation}
\label{EQ:NEWKLBOUND}
    \underline{c}_{\nu}\cdot\ln\left(\frac{\underline{c}_{\nu}+\sqrt{\kappa^{2}+\overline{c}_{\nu}^{2}}}{\underline{c}_{\nu}+\overline{c}_{\nu}}\right)+1
    \leq
    \underline{c}_{\nu}\cdot\ln\left(\kappa+1\right)+1,
\end{equation}
which is at least one due to the non-negativity of $\kappa$.

Substituting \cref{EQ:SUBA1,EQ:NEWKLBOUND} into the infimand of \cref{EQ:APPLY-BEGIN'S-THEOREM}, we obtain
\begin{equation}
\label{EQ:INFIMAND}
    4c\cdot\sqrt{\frac{\nu+1}{\kappa}}+\frac{a\left(\lambda,n\right)}{\lambda}+\frac{\underline{c}_{\nu}}{\lambda}\cdot\ln\left(\kappa+1\right)+\frac{1}{\lambda}\cdot\ln\left(\frac{2e}{\epsilon}\right)+\sqrt{\frac{1}{2n}\cdot\left(\underline{c}_{\nu}\cdot\ln\left(\kappa+1\right)+\ln\left(\frac{4e\cdot\sqrt{n}}{\epsilon}\right)\right)},
\end{equation}
which we emphasise is an upper bound on the infimum.
Our objective is to minimise \cref{EQ:INFIMAND} by appropriately choosing $\kappa$ and $\lambda$ alongside the functional form of $a$.
Accordingly, we let $\kappa=n$ and $\lambda=\sqrt{n}$ alongside $a\left(\lambda,n\right)/\lambda=1/\lambda$.
Given these choices, it then follows that \cref{EQ:INFIMAND} is itself bounded from above by 
\begin{equation}
   4c\cdot\sqrt{\frac{\nu+1}{n}\vphantom{\frac{1}{2n}\ln\left(\frac{4e\cdot\sqrt{n}}{\epsilon}\right)}}+\frac{1}{\sqrt{n}}+\frac{\underline{c}_{\nu}}{\sqrt{n}}\cdot\ln\left(n+1\right)+\frac{1}{\sqrt{n}}\cdot\ln\left(\frac{2e}{\epsilon}\right)+\sqrt{\frac{\underline{c}_{\nu}}{2n}\cdot\ln\left(n+1\right)\vphantom{\frac{1}{2n}\ln\left(\frac{4e\cdot\sqrt{n}}{\epsilon}\right)}}+\sqrt{\frac{1}{2n}\ln\left(\frac{4e\cdot\sqrt{n}}{\epsilon}\right)}
\end{equation}
which exploits the concavity of the square root.
Defining 
\begin{equation}
    M
    \equiv
    4c\cdot\sqrt{\nu+1}+1+\underline{c}_{\nu}\cdot\frac{\ln\left(9\right)}{\ln\left(8\right)}+\ln\left(\frac{2e}{\epsilon}\right)+\sqrt{\frac{\underline{c}_{\nu}\cdot\ln\left(9\right)}{2\cdot\ln\left(8\right)}}+\sqrt{\frac{1}{2}\cdot\ln\left(\frac{4e}{\epsilon}\right)}+\frac{\sqrt{\ln\left(8\right)}}{\sqrt{2}\cdot\ln\left(8\right)},
\end{equation}
which relies on the maintained assumption of \cref{THM:RATE} that $n\geq 8$, we then determine that
\begin{equation}
    R^{\tilde{\Pi}}
    \leq
    \bar{R}+M\cdot\frac{\ln\left(n\right)}{\sqrt{n}},
\end{equation}
such that $M$ is a universal constant.


\subsection{Proof of \cref{THM:CIRCULAR}}
\label{PROOF:CIRCULAR}

The circular variance of a von Mises-Fisher random vector is one minus the ratio of consecutive modified Bessel functions of the first kind (i.e., one minus the mean resultant length).
Bounds on these ratios are derived in \citeauthor{amos1974computation} (\citeyear{amos1974computation}; restated for immediate application to hyperspherical problems in \citealp{KitagawaRowley}).
The result then immediately follows upon subtraction of the lower bound from one.


\subsection{Proof of \cref{THM:KL-RATE}}
\label{PROOF:KL-RATE}

\citet{KitagawaRowley} shows that, when $\Pi$ is a von Mises-Fisher distribution and $\pi_{0}$ is the uniform distribution over the hypersphere,
\begin{equation}
\label{EQ:KLExpansion}
\mathrm{KL}\left(\Pi\|\pi_{0}\right)
=
\nu\cdot\ln\left(\frac{\kappa}{2}\right)-\underbrace{\ln\left(I_{\nu}\left(\kappa\right)\right)}_{\text{Bessel fn.}}-\ln\left(\Gamma\left(\nu+1\right)\right)+\underbrace{A_{m}\left(\kappa\right)\cdot\kappa}_{\text{Ratio fn.}},
\end{equation}
where we recall that $\nu=m/2-1$, with $\underline{c}_{\nu}=\nu+1/2$ and $\overline{c}_{\nu}=\nu+3/2$.
To derive an upper bound on the Kullback-Leibler divergence that does not involve modified Bessel functions or their ratios, we replace the terms labelled \textit{Bessel fn.} and \textit{Ratio fn.} in \cref{EQ:KLExpansion} with appropriate lower and upper bounds, respectively.
To do so, we rely on results (and notation) in \citet{KitagawaRowley}, which are themselves adapted and restated from \citet{amos1974computation}.

First, the term labelled \textit{Bessel fn.} is bounded from below (recall that the term enters negatively) by
\begin{equation}
\label{EQ:BESSELFN.}
\frac{1}{2}\cdot\ln\left(\frac{2}{\kappa}\right)-\ln\left(\Gamma\left(\nu+1\right)\right)+\underline{c}_{\nu}\cdot\ln\left(\frac{\kappa\cdot\left(\nu+1\right)}{\underline{c}_{\nu}+\sqrt{\kappa^{2}+\overline{c}_{\nu}^{2}}}\right)+\frac{\kappa^{2}}{\overline{c}_{\nu}+\sqrt{\kappa^{2}+\overline{c}_{\nu}^{2}}}.
\end{equation}
Second, the term labelled \textit{Ratio fn.} is bounded from above by
\begin{equation}
\label{EQ:RATIOFN.}
    \frac{\kappa^{2}}{\underline{c}_{\nu}+\sqrt{\kappa^{2}+\underline{c}_{\nu}^{2}}}.
\end{equation}
Substituting \cref{EQ:BESSELFN.,EQ:RATIOFN.} into \cref{EQ:KLExpansion}, cancelling terms and noting that
\begin{equation}
    \nu\cdot\ln\left(\frac{\kappa}{2}\right)-\frac{1}{2}\cdot\ln\left(\frac{2}{\kappa}\right)
    =
    \left(\nu+\frac{1}{2}\right)\cdot\ln\left(\frac{\kappa}{2}\right),
\end{equation}
we obtain 
\begin{equation}
\label{EQ:KLUNSIMPLIFIED}
    \underline{c}_{\nu}\cdot\ln\left(\frac{\kappa}{2}\right)-\underline{c}_{\nu}\cdot\ln\left(\frac{\kappa\cdot\left(\nu+1\right)}{\underline{c}_{\nu}+\sqrt{\kappa^{2}+\overline{c}_{\nu}^{2}}}\right)-\frac{\kappa^{2}}{\overline{c}_{\nu}+\sqrt{\kappa^{2}+\overline{c}_{\nu}^{2}}}+\frac{\kappa^{2}}{\underline{c}_{\nu}+\sqrt{\kappa^{2}+\underline{c}_{\nu}^{2}}}.
\end{equation}
To obtain the required result, we simplify the first two terms of \cref{EQ:KLUNSIMPLIFIED} and use the fact that, for any real $c$ (in this case, equal to either $\underline{c}_{\nu}$ or $\overline{c}_{\nu}$, which are positive reals),
\begin{equation}
    \frac{\kappa^{2}}{c+\sqrt{\kappa^{2}+c^{2}}}
    =
    \sqrt{\kappa^{2}+c^{2}}-c,
\end{equation}
which makes use of the formula for the difference of two squares.

\begin{singlespace}
\bibliographystyle{ecta}  
\bibliography{references}
\end{singlespace}


\processdelayedfloats

\title{Supplement to \textit{Stochastic treatment choice with empirical welfare updating}}
\date{\today}
\author{
        Toru Kitagawa
            \and
        Hugo Lopez
            \and
        Jeff Rowley
}
\maketitle
\begin{abstract}
    We apply the methods that are outlined in \textit{Stochastic treatment choice with empirical welfare updating} to experimental data from the Job Training Partnership Act Study and to several numerical simulations.
\end{abstract}


\section{Accounting for the cost of treatment in the JTPA Study sample}
\label{SEC:EXTRAJPTA}

In what follows, we distinguish between the JTPA Study sample, which we sometimes refer to as the raw data, and the cost-adjusted JTPA Study sample, which we sometimes refer to as the costed data.
The costed data subtracts the cost of treatment from the outcome of interest (post-programme earnings) of all treated individuals.\footnote{%
We observe that this leads individuals with zero post-programme earnings to have a negative outcome.
This violates Assumption 1, which requires that all outcomes be bounded and non-negative.
We, nonetheless, proceed with this adjustment of post-programme earnings as is.
An alternative would be to add \$774 to all post-programme earnings before subtracting the assumed cost of treatment.}
Following Kitagawa \& Tetenov (2018), we assume that the cost of treatment is \$774. 
We then proceed to search for the optimal stochastic assignment rule, applying the same grid search approach as we outlined in the main text for the raw data, using the costed data.

\begin{figure}[htb]
    \centering
    \caption{Variation in treatment propensity across individuals in the cost-adjusted JTPA Study sample}
    \label{AFIG:PS}
    \includegraphics[width=\textwidth,height=3.5in]{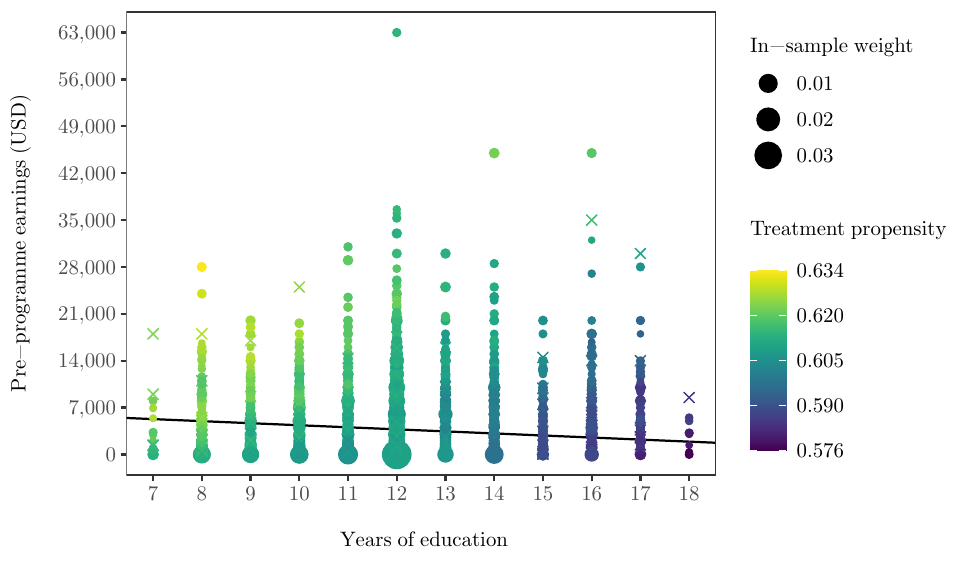}
    \caption*{\footnotesize\textit{Notes:} 
    Cost-adjusted JTPA Study sample.
    This figure illustrates the treatment propensity of individuals under the posterior assignment rule that is induced by $\left(\kappa^{a},\mu^{a}\right)$.
    Each point represents the individual characteristics of an individual or several individuals (crosses denote individuals with negative in-sample weight).
    For comparison, individuals to the left of the solid diagonal line are assigned treatment under the optimal deterministic assignment rule of \citet{Kitagawa2018a}.} 
\end{figure}

\begin{figure}[htb]
    \centering
    \caption{Behaviour of the objective function at $\mu^{*}$ given variation in $\kappa$}
    \label{AFIG:KAPPA}
    \includegraphics[width=\textwidth,height=3.5in]{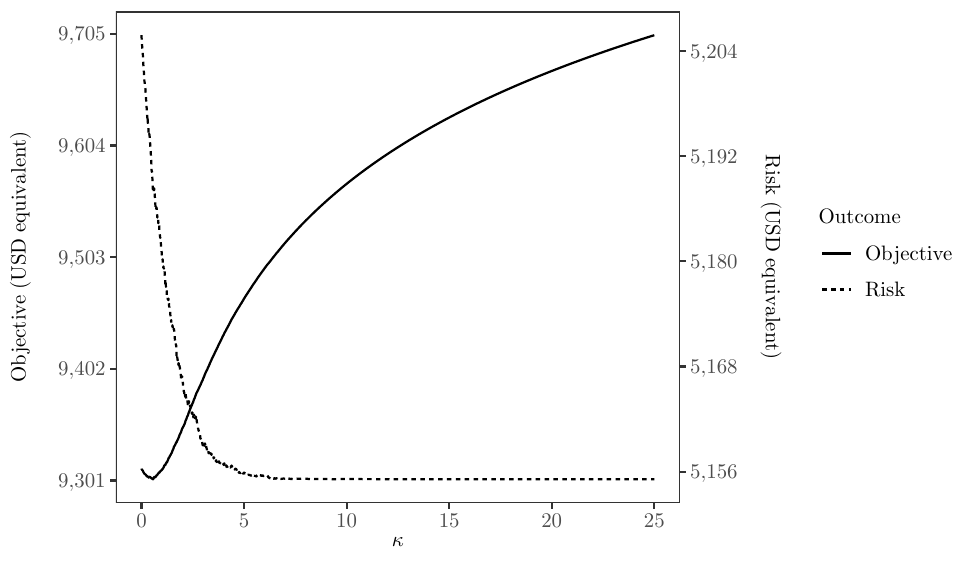}
    \caption*{\footnotesize\textit{Notes:}  
    Cost-adjusted JTPA Study sample.
    This figure illustrates the shape of the objective function and its risk component at $\mu^{a}$ as $\kappa$ is varied.}
\end{figure}

\begin{figure}[htb]
    \centering
    \caption{Deterministic assignment rules and empirical welfare risk}
    \label{AFIG:DETERMINISTICRULES}
    \includegraphics[width=\textwidth,height=3.5in]{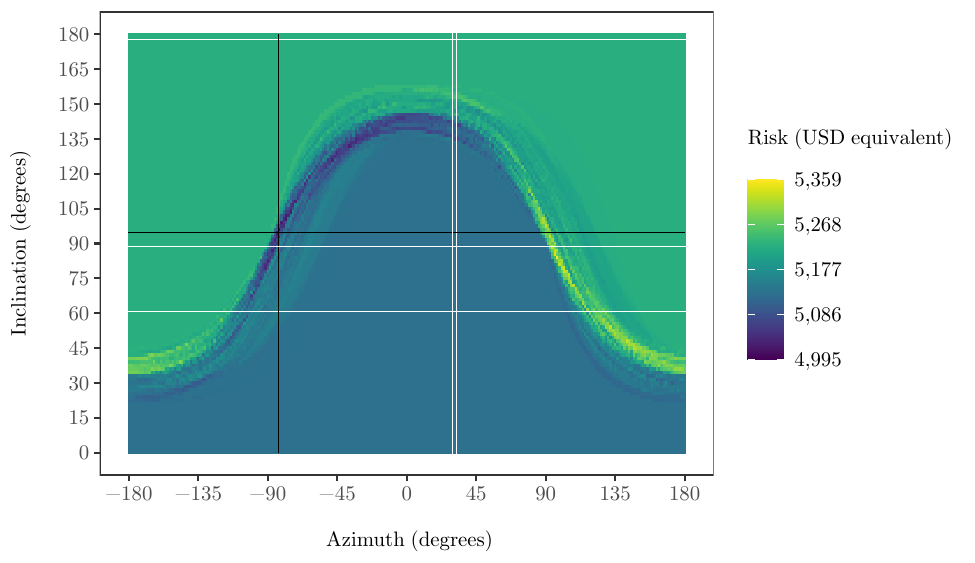}
    \caption*{\footnotesize\textit{Notes:} 
    Cost-adjusted JTPA Study sample.
    This figure illustrates the risk that is associated with (deterministic) assignment rules in $\mathscr{F}$.
    A spherical coordinate mapping is implemented.
    The intersection of the two white lines is located at $\mu^{a}$. 
    The intersection of the two black lines is located at the optimal deterministic assignment rule of \citet{Kitagawa2018a}, which attains the minimal regret amongst all deterministic linear rules.}
\end{figure}

\begin{figure}[htb]
    \centering
    \caption{Behaviour of the objective function at $\kappa^{a}$ given variation in $\mu$}
    \label{AFIG:MU}
    \includegraphics[width=\textwidth,height=3.5in]{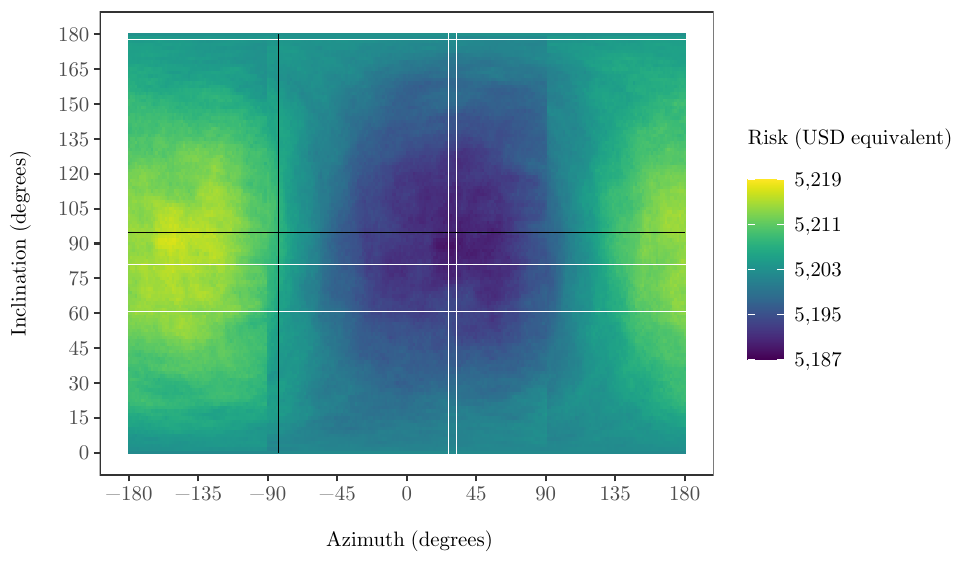}
    \caption*{\footnotesize\textit{Notes:} 
    Cost-adjusted JTPA Study sample.
    This figure illustrates the risk that is associated with (stochastic) assignment rules in $\mathscr{V}$;
    the concentration parameter is fixed at $\kappa^{a}$ whilst $\mu$ is varied.
    The intersection of the two white lines is located at $\mu^{a}$. 
    The intersection of the two black lines is located at the optimal deterministic assignment rule of \citet{Kitagawa2018a}, which attains the minimal regret amongst all deterministic linear rules.}
\end{figure}

\begin{figure}[htb]
    \centering
    \caption{Distribution of propensity of treatment across individuals (adjusted) at $\left(\kappa^{a},\mu^{a}\right)$}
    \label{AFIG:DP-C}
    \includegraphics[width=\textwidth,height=3.5in]{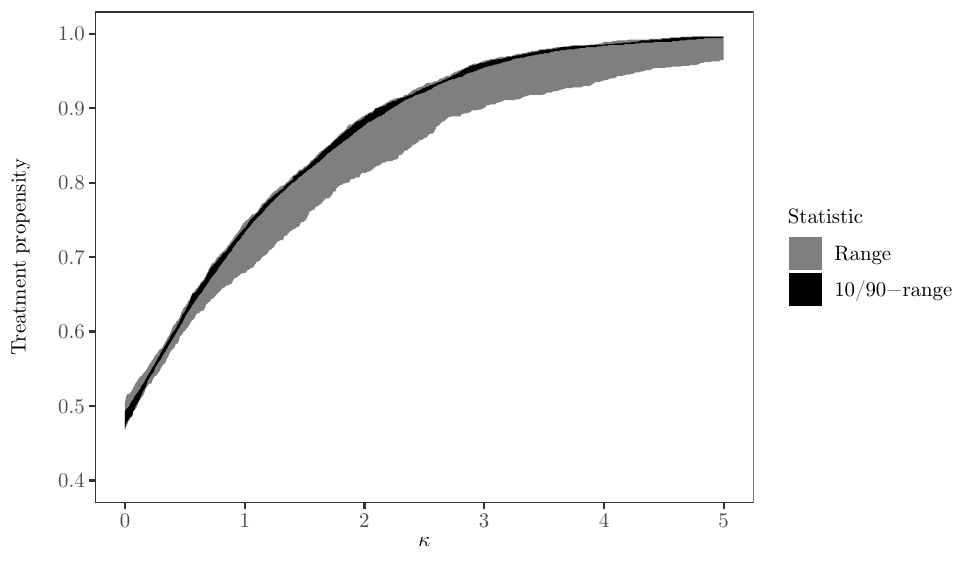}
    \caption*{\footnotesize\textit{Notes:} 
    Cost-adjusted JTPA Study sample.
    This figure illustrates the variation in treatment propensity across individuals.
    We draw 1,000 directional vectors and count how many of these vectors, implemented as deterministic assignment rules, assign each individual to treatment.
    We sort individuals by how often they are assigned to treatment and summarise the implied distribution by the range and the 10\textsuperscript{th} and 90\textsuperscript{th} quantiles.}
\end{figure}

\begin{figure}[htb]
    \centering
    \caption{Distribution of propensity of treatment across individuals (raw) at $\left(\kappa^{*},\mu^{*}\right)$}
    \label{AFIG:DP}
    \includegraphics[width=\textwidth,height=3.5in]{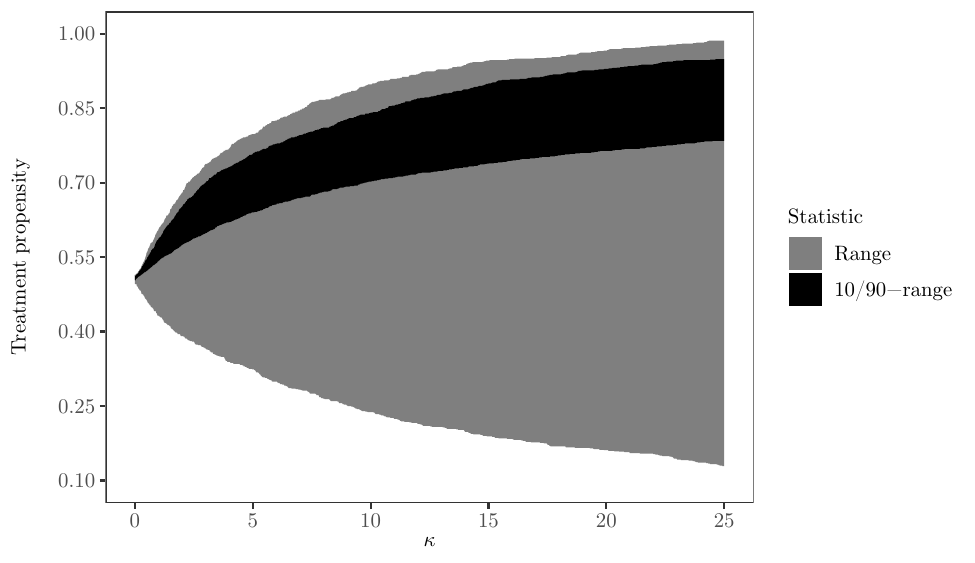}
    \caption*{\footnotesize\textit{Notes:} 
    JTPA Study sample.
    This figure illustrates the variation in treatment propensity across individuals.
    We draw 1,000 directional vectors and count how many of these vectors, implemented as deterministic assignment rules, assign each individual to treatment.
    We sort individuals by how often they are assigned to treatment and summarise the implied distribution by the range and the 10\textsuperscript{th} and 90\textsuperscript{th} quantiles.}
\end{figure}

We find that the objective function is minimised by the stochastic assignment rule with $\kappa=0.560$ and $\mu=(0.872,0.490,0.018)$, which we label $\kappa^{a}$ and $\mu^{a}$, respectively.\footnote{%
This directional vector can be represented by an azimuth of $29^\circ$ and an inclination of $89^\circ$ using spherical coordinates.}
The value of the objective function and the empirical welfare risk induced by this assignment rule are equivalent to \$9,302 and \$5,155, respectively.

In comparison, Kitgagawa \& Tetanov (2018) estimates that the deterministic assignment rule defined by $\beta=(0.117,-0.990,-0.086)$ minimises empirical welfare risk.
Relative to the raw analysis, this assignment rule is much flatter, and essentially determines that all individuals with low pre-programme earnings (earnings of around \$5,000 or less) should be treated, irrespective of education.

\cref{AFIG:PS,AFIG:KAPPA,AFIG:DETERMINISTICRULES,AFIG:MU} are intended to be comparable to the figures in the main text. 
There are two differences of note between the results of the raw and costed analyses. 
First, in \cref{AFIG:KAPPA}, the relationship between $\kappa$ and empirical welfare risk is much more muted, with risk only decreasing by around \$50 as $\kappa$ increases (over the interval that we study). 
In contrast, empirical welfare risk decreases by around \$200 when the cost of treatment is not accounted for.
This reflects the second difference, which is apparent in \cref{AFIG:DETERMINISTICRULES}.
Specifically, that the difference in empirical welfare risk between those assignment rules that assign everyone to treatment versus those that assign no-one to treatment is much smaller.
The \textit{benefit} to an increase in $\kappa$, which is a reduction in the probability mass allocated to those assignment rules that assign no-one to treatment is, accordingly, smaller.
The implication is that the penalty term starts to dominate the objective function for smaller values of $\kappa$.

To supplement our analysis, we include two additional figures in \cref{AFIG:DP-C,AFIG:DP}.
These figures are intended to give some idea about how the concentration of the von Mises-Fisher distribution influences the propensity with which individuals are assigned to the experimental group.
For example, given a particular instance of the von Mises-Fisher distribution, one individual might have a high propensity of assignment whereas another might have a low propensity.
In other words, the number of directional vectors that are drawn from the von Mises-Fisher distribution for which the first individual is assigned to the experimental group is greater than for the second individual.
As $\kappa$ increases, so these directional vectors concentrate around the mean direction, and the propensity of assignment to the experimental group approaches zero or one for each individual. 
We see from \cref{AFIG:DP-C} that, when $\mu$ is fixed at $\mu^*$, the propensity of assignment in the costed data tends towards one for all individuals.
In contrast, and to illustrate the possibility of complete dichotomy, for the raw data, we see from \cref{AFIG:DP} that the propensity of assignment diverges as $\kappa$ increases.
We observe in both datasets that much of the variation in propensity is driven by a few individuals in the tail, who are likely those individuals whose individual characteristics make them outliers.
In both cases, the median propensity of treatment (i.e., the probability with which the average individual in the sample is assigned to the experimental group) tends towards one.


\section{Numerical simulations}
\label{SEC:SIMULATIONS}

We propose several experiments that investigate how various aspects of the sample data that is available to the social planner affect the posterior distribution and its shape when the specified prior distribution is uniform over the sphere.
We conduct these experiments to better understand some of the empirical results that we obtain for the main paper. 

Aspects of the sample data that we vary include the number of observations, the outcome of interest, and individual characteristics.
We conduct these experiments with simulated data and through manipulation of the existing empirical application.
Each experiment is comparable to the existing empirical application in any case, in that observed individual characteristics are taken to be education and pre-programme earnings alongside an intercept term, and the outcome of interest is taken to be post-programme earnings.
We investigate how these aspects and our variation of them affect the objective function.

The specific questions that we ask, and that inform the design of our experiments, are as follows.
\begin{itemize}
\item[--] How does the number of observations influence the shape of the objective function?
\item[--] How does the relative influence of education and pre-programme earnings on post-programme earnings influence empirical welfare risk?
\item[--] How does the distribution of individual characteristics influence the shape of the objective function?
\end{itemize}
To address these questions we propose a series of linear specifications that satisfy the bounded outcomes assumption (i.e., the outcome of interest can be restricted to some subset $[0,M]$, where $M<\infty$) that we require. 
We concede that these specifications are somewhat contrived, reflecting our need to balance tractability with the requirements of the bounded outcomes assumption.
Denoting individual characteristics by ${X}$, we suppose that
\begin{equation}
\label{EQ:X}
{X}=(1,X_\textrm{earn},X_\textrm{educ})\in\{1\}\times[0,1]^{2},
\end{equation}
which can always be maintained via an appropriate affine map of the individual characteristics (a step that we undertake in any case), and propose that
\begin{equation}
\label{EQ:PO}
\arraycolsep=0pt
\begin{array}{lclcl}
Y_{1}
&{}={}&
{X}^{\top}{\eta}+\epsilon_1,&{}\hspace{15pt}{}&\epsilon_1\sim\mathrm{N}(0,\sigma_{1}^{2})\mathrm{\,truncated\,to\,}(0,m),\\
Y_{0}
&{}={}&
{X}^{\top}{\alpha}+\epsilon_0,&{}\hspace{15pt}{}&\epsilon_0\sim\mathrm{N}(0,\sigma_{0}^{2})\mathrm{\,truncated\,to\,}(0,m),\\
\end{array}
\end{equation}
with
\begin{equation}
\arraycolsep=0pt
\begin{array}{lcl}
{\eta}&{}\in{}&\{v\in\mathbb{R}^3\,:\,v_{1}+v_{2}+v_{3}\geq0\},\\
{\alpha}&{}\in{}&\{v\in\mathbb{R}^3\,:\,v_{1}+v_{2}+v_{3}\geq0\},
\end{array}
\end{equation}
where $Y_{1}$ and $Y_{0}$ denote the potential outcomes associated with treatment and non-treatment, respectively. 
The advantage of this specification is that it provides a clear interpretation of the influence of education and pre-programme earnings on post-programme earnings, and facilitates addressing all of the questions that we pose.
Moreover, the potential outcomes are restricted to the lie in the union of the sets  $[0,\eta+m]$ and $[0,\alpha+m]$.
We note that the conditional expectations of the potential outcomes are
\begin{equation}
\label{EQ:CE}
\arraycolsep=0pt
\begin{array}{lcl}
\mathrm{E}(Y_{1}|{x})
&{}={}&
{x}^{\top}{\eta}+\frac{\phi(0)-\phi(m/\sigma_{1})}{\Phi(m/\sigma_{1})-\Phi(0)}\cdot\sigma_{1}
=
{x}^{\top}\tilde{{\eta}},\\
\mathrm{E}(Y_{0}|{x})
&{}={}&
{x}^{\top}{\alpha}+\frac{\phi(0)-\phi(m/\sigma_{0})}{\Phi(m/\sigma_{0})-\Phi(0)}\cdot\sigma_{0}
=
{x}^{\top}\tilde{{\alpha}},
\end{array}
\end{equation}
which are both affine functions of individual characteristics.\footnote{%
Other commonly invoked models fail to meet our test of tractability or the requirements of the bounded outcomes assumption: the standard censored outcome model generates a rectified normal distribution that is non-linear in individual characteristics and is unbounded from above; logarithmic transformation of the outcome implies a non-linear transformation of individual characteristics.}
This is an attractive property that we exploit.

Knowledge of the data generating process (\cref{EQ:PO} in our framework) is sufficient to determine the optimal assignment rule.\footnote{%
We do not make any claims about uniqueness in what follows and, indeed, presented with a finite sample of individual characteristics, it is likely that several assignment rules can attain the same partition of individuals as what we refer to as the optimal assignment rule.}
Specifically, the optimal assignment rule delivers the partition
\begin{equation}
G
=
\left\{{x}\,:\,\mathrm{E}(Y_{1}|{x})\geq\mathrm{E}(Y_{0}|{x})\right\}.
\end{equation}
When the conditional expectation of the potential outcomes are affine functions of individual characteristics (and assignment is at random, which is an assumption that we implicitly maintain throughout) then the optimal assignment rule belongs to the LES class.
Hence, \cref{EQ:CE} guarantees that the optimal policy has the specific form
\begin{equation}
\label{EQ:OTR}
G
=
\{{x}\,:\,{x}^{\top}{\beta}\geq0,{\beta}=(\tilde{{\eta}}-\tilde{{\alpha}})/\|\tilde{{\eta}}-\tilde{{\alpha}}\|_2\},
\end{equation}
which is not only a member of the LES class but emphasises that the optimal assignment rule can be summarised by a vector on the sphere (provided that the potential outcomes have distinct process).
This result provides motivation for our reliance on spherical distributions and, in particular, the von Mises-Fisher distribution.
We observe that, abstracting from the issue of sampling variation, empirical welfare risk is minimised when the mean direction of the von Mises-Fisher distribution coincides with the policy defined in \cref{EQ:OTR}.
This insight allows us to focus exclusively on the influence of the concentration parameter on empirical welfare risk and the objective function.

Throughout, we are careful to sample using inversion-based pseudo-random sampling methods where possible. 
Compared to rejection sampling-based pseudo-random sampling methods, inversion-based methods are able to guarantee comparability across experiments despite differences in parameter values. 
We also note that where we use data from the JTPA Study sample, this data is not adjusted for the cost of treatment (i.e., we use the raw data).
We continue to label those values of the parameters that minimise the objective function by $\kappa^{*}$ and $\mu^{*}$, which we emphasise can vary across the various experiments that we conduct.


\subsection{Varying the number of observations}

\begin{figure}[htb]
    \centering
    \caption{Behaviour of the objective function at $\mu^{*}$ given variation in $\kappa$}
    \label{AFIG:KAPPAN}
    \includegraphics[width=\textwidth,height=2.25in]{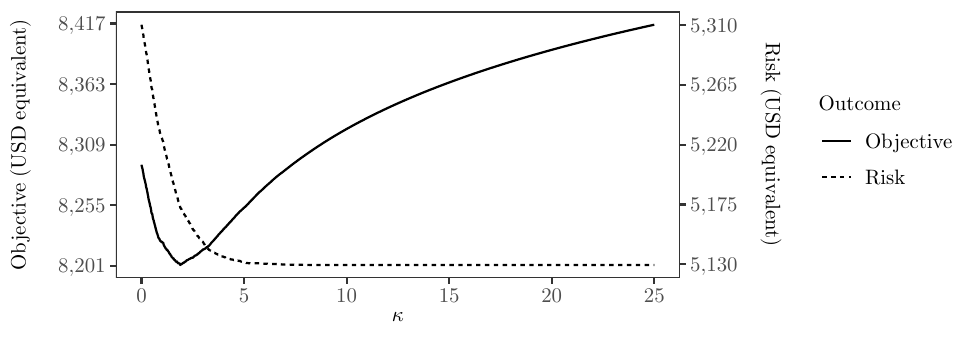}
    \includegraphics[width=\textwidth,height=2.25in]{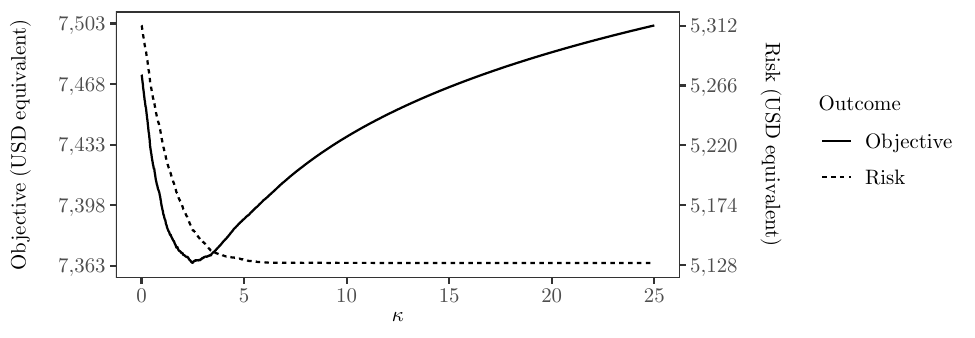}
    \caption*{\footnotesize\textit{Notes:}
    JTPA Study sample (artificially inflated by copying observations).
    This figure illustrates how the objective function and its risk component change as the number of observations is doubled (upper panel) and quadrupled (lower panel).
    We hold $\mu$ fixed at $\mu^{*}$ in each case, which differs according to the sample size, and vary $\kappa$.}
\end{figure}

To investigate the influence that the number of observations has on the objective function and on empirical welfare risk, we copy each observation in the JTPA Study sample either once or three times, thereby doubling and quadrupling, the number of individuals in the sample.
The mechanical effect of this change is to mute the influence of the penalty term in the objective function, without affecting empirical welfare risk.
The immediate implication is that larger values of the concentration parameter can be sustained, since the penalty due to deviating from the uniform distribution is relatively smaller for every such deviation.

We plot the objective function and empirical welfare risk for the doubled and quadrupled JTPA Study samples in \cref{AFIG:KAPPAN}. 
We observe that increasing the number of observations has the direct effect of decreasing the magnitude of the objective function.
Moreover, increasing the number of observations also leads to an increase in $\kappa^{*}$, which increases from 1.550 to 1.890 and then to 2.490.
We emphasise that $\mu^{*}$ is also not the same across the two cases.
The intuition here is that an increase in $\kappa$ leads to the concentric contour map of the density function becoming more tightly arranged around $\mu$ (whatever that may be).
Locating $\mu$ closer to the boundary between high and moderate regret regions say, such as where the deterministic assignment rule of Kitagawa \& Tetenov (2018) is located, incurs less of a penalty in this instance since the density function assigns less probability mass to the high regret region than it would for a smaller value of the $\kappa$.
For the doubled sample we find that $\mu^{*}=(0.812,0.577,0.088)$, whilst for the quadrupled sample we find that $\mu^{*}=(0.917,0.394,0.053)$.\footnote{%
These directions translate (azimuth:inclination) to $35^{\circ}:85^{\circ}$ and $23^{\circ}:87^{\circ}$, respectively, as compared to $\mu^{*}=(0.883,0.442,0.158)$ or $27^{\circ}:81^{\circ}$ in the original sample.}


\subsection{The variance of post-programme earnings}

In the following three experiments, we investigate the influence of the variance of post-programme earnings on the results of our method.
Specifically, we investigate how altering $\sigma_{0}$ and $\sigma_{1}$ in the specification outlined in \cref{EQ:PO} affects our results.
In particular, we are interested in whether there is a fundamental change in how empirical welfare risk varies with the parameters of the von Mises-Fisher distribution.

\begin{figure}[htb]
    \centering
    \caption{(Experiment 1) Variation in treatment propensity across individuals}
    \label{AFIG:PS-1}
    \includegraphics[width=\textwidth,height=3.5in]{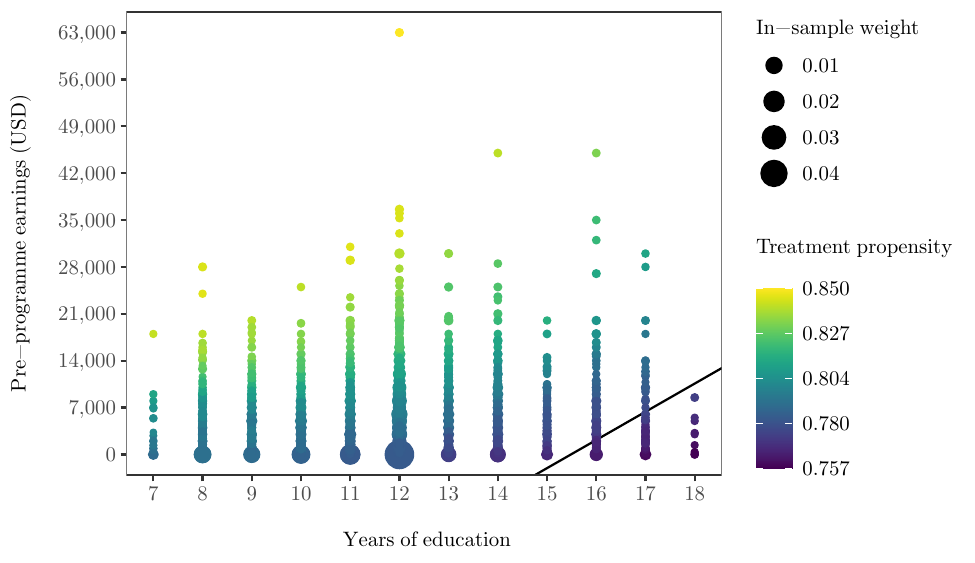}
    \caption*{\footnotesize\textit{Notes:}
    Simulated data generated using \cref{EQ:BASELINE-1} in conjunction with data from the JTPA Study sample.
    This figure illustrates the treatment propensity of individuals in a simulated sample under the posterior assignment rule that is induced by $\left(\kappa^{*},\mu^{*}\right)$.
    Each point represents the individual characteristics of an individual or several individuals.
    For comparison, individuals to the left of the solid diagonal line are assigned treatment under the oracle assignment rule of \cref{EQ:EXP-1-OPTIMAL}.} 
\end{figure}

The JTPA Study sample consists of 9,223 observations, and we extract treatment status and individual characteristics directly from this dataset.
We then generate potential outcomes according to \cref{EQ:PO} (\cpageref{EQ:PO}).
We implement the affine map of \cref{EQ:X} via the transformations
\begin{equation}
\arraycolsep=0pt
\begin{array}{lcl}
X_\textrm{earn}&{}\mapsto{}&X_\textrm{earn}/\max\{X_\textrm{earn}\},\\
X_\textrm{educ}&{}\mapsto{}&X_\textrm{educ}/\max\{X_\textrm{educ}\}.\\
\end{array}
\end{equation}
We then regress post-programme earnings in the experimental group and the control group on these characteristics, separately and together, so that the baseline experiment somewhat mimics the JTPA Study sample.
Using our simple regressions as a rough guide, we let
\begin{equation}
\label{EQ:BASELINE-1}
\arraycolsep=0pt
\begin{array}{lclclcl}
{\eta}&{}={}&(+3,040,+86,446,+14,008),&{}\hspace{15pt}{}&\sigma_{1}&{}={}&15,914,\\
{\alpha}&{}={}&(-1,086,+82,458,+18,804),&{}\hspace{15pt}{}&\sigma_{0}&{}={}&15,914,
\end{array}
\hspace{15pt}m=5\sigma_{1}=5\sigma_{0},
\end{equation}
so that the optimal policy is 
\begin{equation}
G
=
\{{x}\,:\,{x}^{\top}{\beta}\geq0,{\beta}=(+0.552,+0.533,-0.641)\},
\label{EQ:EXP-1-OPTIMAL}
\end{equation}
which we plot in \cref{AFIG:PS-1}.
We note that the coefficients of the linear specification are so large because our simple regressions scale individual characteristics but do not scale post-programme earnings, which inflates the effect of pre-programme earnings and years of education.
This constitutes our first experiment.

We then propose two further experiments. 
Our second experiment increases the variance of post-programme earnings by inflating $\sigma_{1}$, $\sigma_{0}$ and $m$ in \cref{EQ:BASELINE-1} by a factor of five.
Our third experiment similarly reduces the variance of post-programme earnings by deflating $\sigma_{1}$, $\sigma_{0}$ and $m$ in \cref{EQ:BASELINE-1} by a factor of five. 
We note that both experiments leave the oracle policy unchanged from \cref{EQ:EXP-1-OPTIMAL}.
We present the corresponding estimates of the parameters of the posterior distribution immediately below.

\begin{center}
\begin{tabular}{lcc}
         \hline
         \text{Experiment}& $\mu^{*}$ &$\kappa^{*}$\\
         
         \hline
         
         \text{1. Baseline}			& (+0.822,+0.565,+0.078) & 1.850 \\

    \text{2. High variance}  				& (+0.641,+0.703,-0.309) & 0.340 \\

    \text{3. Low variance }	& (+0.831,+0.547,+0.105) & 1.970 \\ 
    
    \hline
\end{tabular}
\end{center}

An immediate conclusion that we can draw from these results is that $\mu^{*}$ does not align with the oracle assignment rule.
That being said, it is apparent from \cref{AFIG:PS-1} that individuals who are assigned treatment under the oracle assignment rule are more likely to be assigned treatment under the posterior distribution that we obtain.

\begin{figure}[htb]
    \centering
    \caption{(Experiments 1--3) Behaviour of the objective function at $\mu^{*}$ given variation in $\kappa$}
    \label{AFIG:KAPPA-123}
    \includegraphics[width=3in,height=5in]{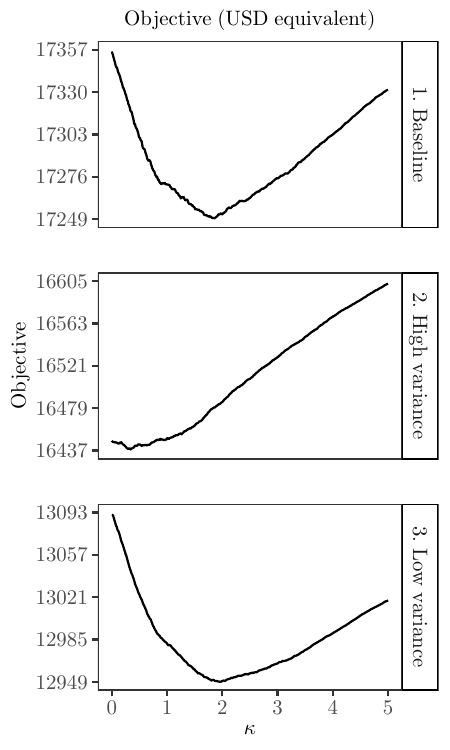}
        \includegraphics[width=3in,height=5in]{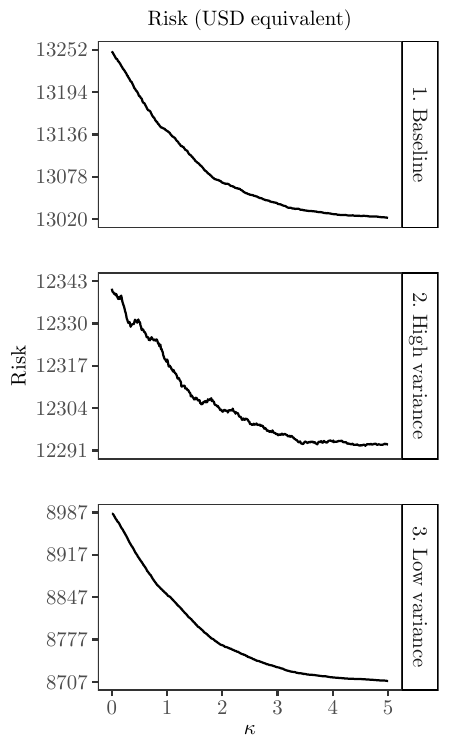}
    \caption*{\footnotesize\textit{Notes:}
    Simulated data generated using \cref{EQ:BASELINE-1} in conjunction with data from the JTPA Study sample.
    This figure illustrates how the shape of the objective function and its risk component changes as the variance of post-programme earnings is increased (middle panel) and decreased (lower panel). 
    We hold $\mu$ fixed at $\mu^{*}$ in each case, which varies according to the sample size, and vary $\kappa$.}
\end{figure}

We plot the behaviour of the objective function and of empirical welfare risk in \cref{AFIG:KAPPA-123}.
We emphasise the non-smoothness of empirical welfare risk for the second experiment, and we suggest that increasing the variance of post-programme earnings makes the problem of finding the optimal assignment rule more difficult.


\subsection{The location of the oracle assignment rule}

A common feature of the data generating process outlined in \cref{EQ:BASELINE-1} for Experiments 1 through 3 is the oracle assignment rule, which is located in the south-east corner of the covariate space (\cref{AFIG:PS-1}).
Two things can be inferred from this.
First, that only a small minority of individuals that are educated to post-graduate level are not assigned treatment under the oracle assignment rule.
In other words, a small subset of the data with extreme characteristics.
Second, that, all else being equal, average outcomes in the experimental group are higher than average outcomes in the control group, and that this baseline difference is relatively important as compared to pre-programme earnings and years of education.
These features, we suggest, are indicative of a relatively easy problem (of determining the optimal assignment rule).

\begin{figure}[htb]
    \centering
    \caption{(Experiment 4) Variation in treatment propensity across individuals}
    \label{AFIG:PS-4}
    \includegraphics[width=\textwidth,height=3.5in]{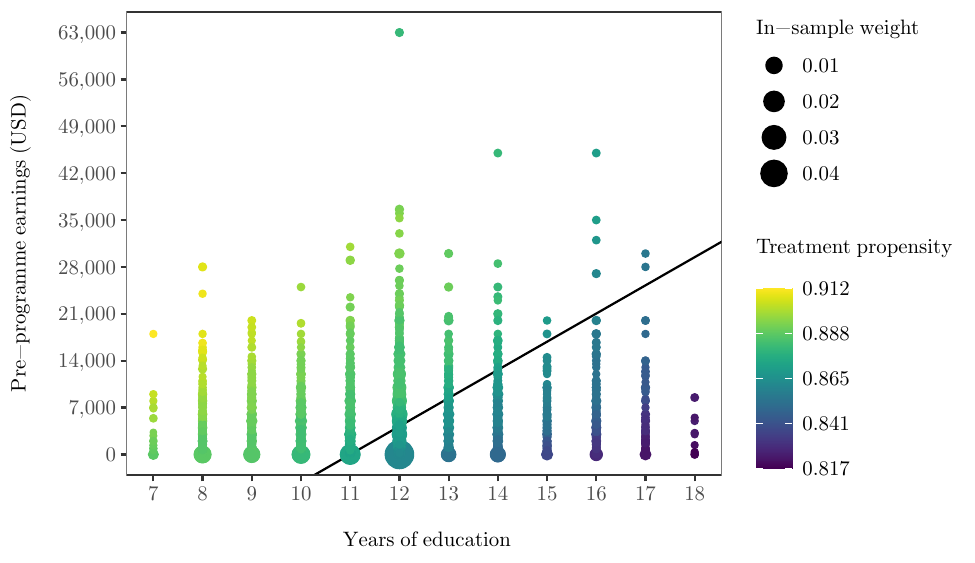}
    \caption*{
    \footnotesize\textit{Notes:}
    Simulated data generated using \cref{EQ:BASELINE-4} in conjunction with data from the JTPA Study sample.
    This figure illustrates the treatment propensity of individuals in a simulated sample under the posterior assignment rule that is induced by $\left(\kappa^{*},\mu^{*}\right)$.
    Each point represents the individual characteristics of an individual or several individuals.
    For comparison, individuals to the left of the solid diagonal line are assigned treatment under the oracle assignment rule of \cref{EQ:EXP-4-OPTIMAL}.} 
\end{figure}

The following three experiments largely replicate Experiments 1 through 3 but shift the oracle assignment rule to the left, narrowing the difference between the baseline average outcomes in the experimental and control groups.
In Experiment 4, we assume that 
\begin{equation}
\label{EQ:BASELINE-4}
\arraycolsep=0pt
\begin{array}{lclclcl}
{\eta}&{}={}&(+2,442,+86,446,+14,008),&{}\hspace{15pt}{}&\sigma_{1}&{}={}&15,914,\\
{\alpha}&{}={}&(-489,+82,458,+18,804),&{}\hspace{15pt}{}&\sigma_{0}&{}={}&15,914,
\end{array}
\hspace{15pt}m=5\sigma_{1}=5\sigma_{0},
\end{equation}
so that the optimal policy is 
\begin{equation}
G
=
\{{x}\,:\,{x}\cdot{\beta}\geq0,{\beta}=(+0.425,+0.579,-0.696)\},
\label{EQ:EXP-4-OPTIMAL}
\end{equation}
which we plot in \cref{AFIG:PS-4}.
Experiments 5 and 6 then mirror Experiments 2 and 3 in that they inflate and deflate, respectively, the variance of post-programme earnings.
We suggest that \cref{EQ:BASELINE-4} is a more difficult problem that \cref{EQ:BASELINE-1}.

We specifically design these experiments so that the oracle assignment rule is such that as close to 50\% of the sample is assigned to treatment as is possible without altering the importance of pre-programme earnings relative to years of education in the earnings process. 
We present the corresponding estimates of the parameters of the posterior distribution immediately below.

\begin{center}
\begin{tabular}{lcc}
         \hline
         \text{Experiment}& $\mu^{*}$ &$\kappa^{*}$\\
         
         \hline
         
         \text{4. Baseline}			& (+0.932,+0.362,-0.035) & 0.00 \\

    \text{5. High variance}  				& (-0.810,-0.526,-0.259) & 0.390 \\

    \text{6. Low variance }	& (+0.899,+0.425,+0.105) & 0.00 \\ 
    
    \hline
\end{tabular}
\end{center}

\begin{figure}[htb]
    \centering
    \caption{(Experiments 4--6) Behaviour of the objective function at $\mu^{*}$ given variation in $\kappa$}
    \label{AFIG:KAPPA-456}
    \includegraphics[width=3in,height=5in]{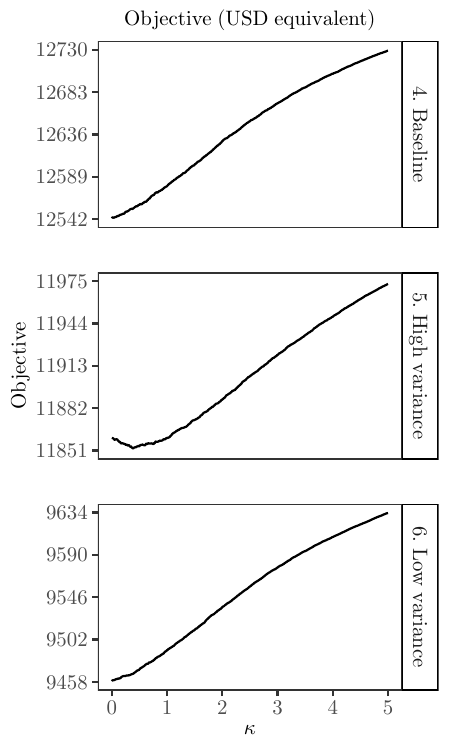}
    \includegraphics[width=3in,height=5in]{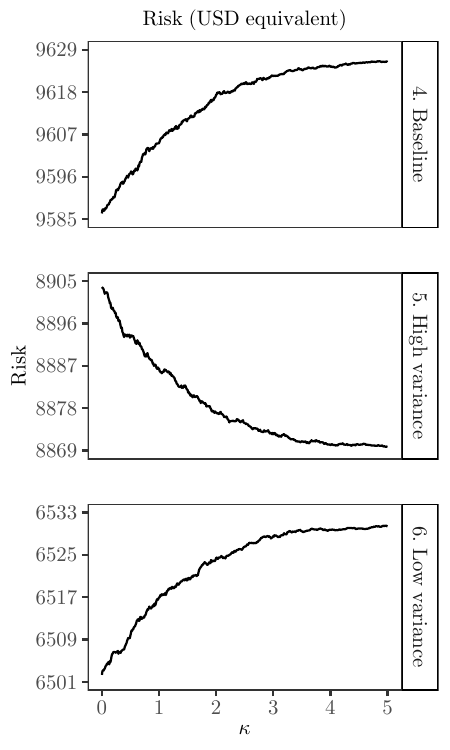}
    \caption*{\footnotesize\textit{Notes:}
    Simulated data generated using \cref{EQ:BASELINE-4} in conjunction with data from the JTPA Study sample.
    This figure illustrates how the shape of the objective function and its risk component changes as the variance of post-programme earnings is increased (middle panel) and decreased (lower panel). 
    We hold $\mu$ fixed at $\mu^{*}$ in each case, which differs according to the sample size, and vary $\kappa$.}
\end{figure}

A curious feature of these results is that, for Experiments 4 and 6, the posterior distribution is uniform.
We note that the mean direction is irrelevant in this case, but we include it anyway because knowledge of it is necessary to interpret \cref{AFIG:KAPPA-456}.
This uniformity of the posterior distribution is apparently driven by how empirical welfare risk increases alongside $\kappa$ in \cref{AFIG:KAPPA-456}.
A possible explanation for this feature is that this data generating process is indeed hard.
We suggest that the narrow gap between the baseline average outcomes of the experimental and control groups makes treating everyone versus treating no-one equally appealing (or unappealing). 
For instance, contrasting \cref{AFIG:DETERMINISTICRULES} with its analogue in the main text (for the raw data), we see that the difference in empirical welfare risk between the \textit{treat everyone rules} and \textit{treat no-one rules} that occupy the southern and northern regions of the heatmaps, respectively, narrows.
A similar effect happens here.
Given the fact that the density contours of the von Mises-Fisher distribution are concentric, there is then no natural region of the sphere to locate in order to minimise empirical welfare risk unless $\kappa$ is particularly large and the posterior distribution allocates substantial probability mass to an extremely localised area on the frontier between the collections of \textit{treat everyone rules} and \textit{treat no-one rules}.
High concentration is heavily penalised by the Kullback-Leibler divergence though, and so uniformity cannot be improved upon.
The fact that a different pattern is observed for Experiment 5 is compatible with this argument, and could be achieved if the high variance of post-programme earnings makes a few observations in the sample pivotal (recall that the weight attached to each observation is proportional to post-programme earnings; inflating the variance of post-programme earnings can lead to greater concentration of weight on a few observations, since extreme outliers are more likely).


\subsection{Individual characteristics and their distribution}

\begin{figure}[htb]
    \centering
    \caption{(Experiments 7--8) Variation in treatment propensity across individuals}
    \label{AFIG:PS-78}
    \includegraphics[width=\textwidth,height=3.5in]{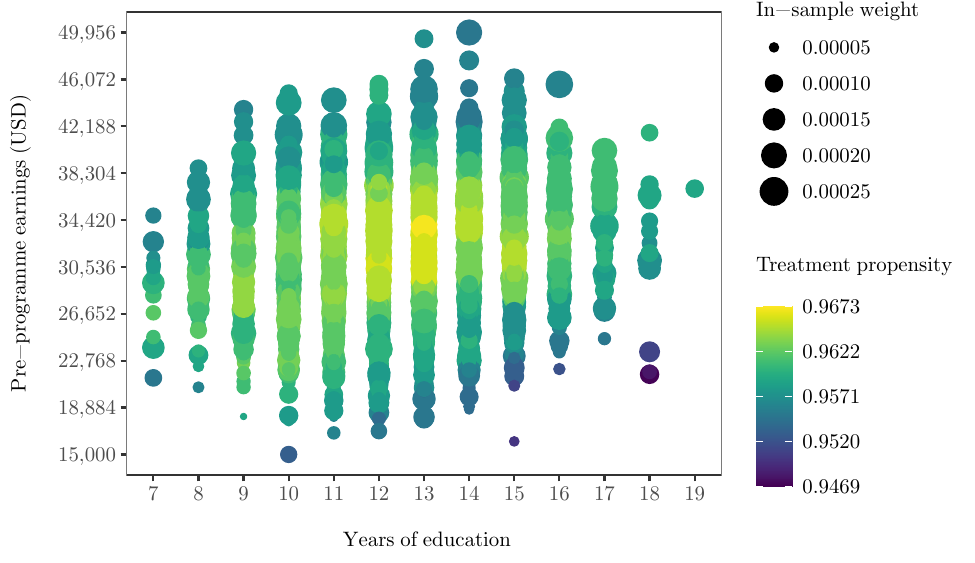}
    \includegraphics[width=\textwidth,height=3.5in]{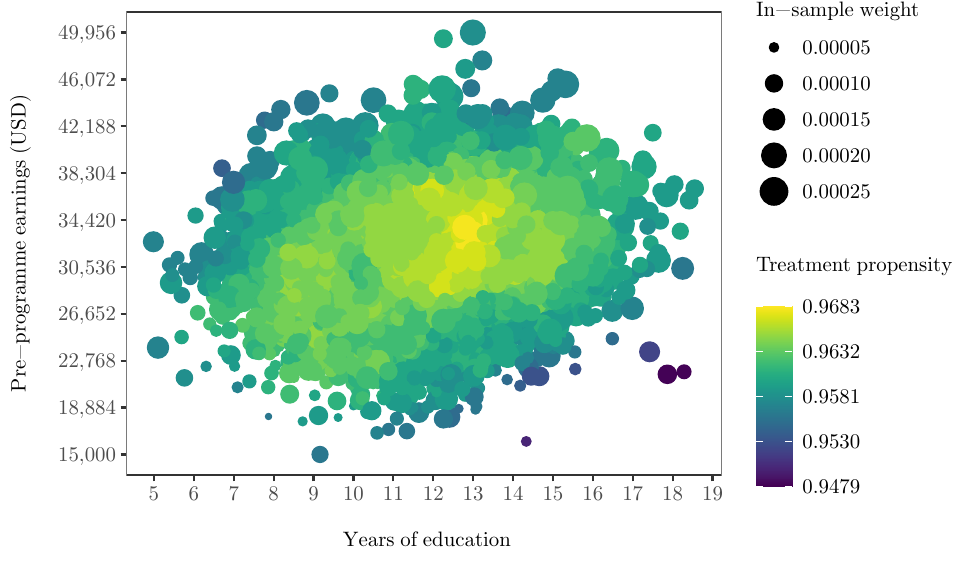}
    \caption*{\footnotesize\textit{Notes:}
    Simulated data generated using \cref{EQ:BASELINE-1} and a bivariate normal distribution.
    This figure illustrates the treatment propensity of individuals in simulated samples under the posterior assignment rule that is induced by $\left(\kappa^{*},\mu^{*}\right)$.
    Each point represents the individual characteristics of an individual or several individuals.
    Every individual is assigned to treatment under the oracle assignment rule.}
\end{figure}

The final experiments that we undertake vary the distribution of individual characteristics.
Our intention is to understand whether the JTPA Study sample is, in some sense, special and whether altering individual characteristics substantially alters our results.

Experiments 7 and 8 mirror Experiment 1 in how the outcome is generated: 
both experiments rely on \cref{EQ:PO} and the parameter values that we outline in \cref{EQ:BASELINE-1} for Experiment 1. 
The distinction between Experiments 7 and 8 and Experiment 1 is how individual characteristics are generated. 
Whereas Experiment 1 uses data taken from the JTPA Study sample, Experiments 7 and 8 generate individual characteristics according to a bivariate normal distribution with the mean vector equal to half of the maximum of pre-programme earnings and the average number of years of education, respectively. 
We estimate the covariance matrix of pre-programme earnings and years of education in the JTPA Study sample, and set the covariance of the bivariate normal distribution equal to this (Pearson correlation coefficient of 0.126).
In the case of Experiment 7, we discretise years of education by assigning each observation to one of 12 equal-sized bins.
We then map each characteristic to the unit interval by means of the aforementioned linear transformation.

\begin{figure}[htb]
    \centering
    \caption{(Experiments 7--8) Behaviour of the objective function at $\mu^{*}$ given variation in $\kappa$}
    \label{AFIG:KAPPA-78}
    \includegraphics[width=3in,height=3.5in]{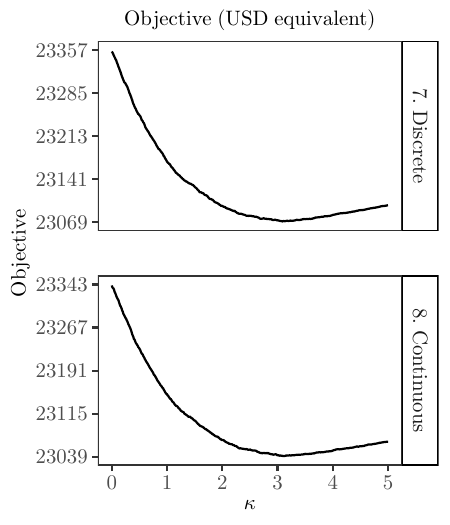}
        \includegraphics[width=3in,height=3.5in]{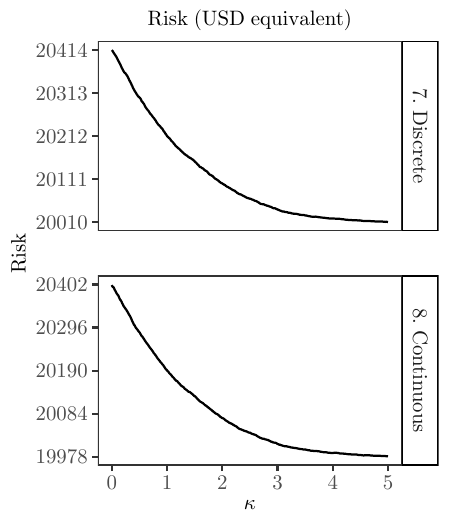}
    \caption*{\footnotesize\textit{Notes:}
    Simulated data generated using \cref{EQ:BASELINE-1} and a bivariate normal distribution.
    This figure illustrates how the shape of the objective function and its risk component at $\mu^{*}$ changes as $\kappa$ increases. }
\end{figure}

We plot the individual characteristics that we use in Experiments 7 and 8 in \cref{AFIG:PS-78}.
Absent from either plot is the oracle assignment rule.
This is not an oversight.
Rather, the oracle assignment rule is such that it recommends that all individuals be assigned treatment (i.e., it to the south-east of the plotting area).
We plot the criterion (left-hand panel) and empirical welfare risk (right-hand panel) for Experiments 7 and 8 in \cref{AFIG:KAPPA-78} (\cpageref{AFIG:KAPPA-78}).
Due to the similarity of \cref{AFIG:KAPPA-78} to the other figures that we have presented, we do not provide any further discussion of this figure.
What is somewhat interesting though is how the propensity of treatment is highest for those individuals with individual characteristics located around their mean values.

We present the corresponding estimates of the parameters of the posterior distribution immediately below.

\begin{center}
\begin{tabular}{lcc}
         \hline
         \text{Experiment}& $\mu^{*}$ &$\kappa^{*}$\\
         
         \hline
         
         \text{7. Discrete}			& (+0.794,+0.405,+0.454) & 3.090 \\

    \text{8. Continuous}  				& (+0.801,+0.408,+0.438) & 3.120 \\
    
    \hline
\end{tabular}
\end{center}

We find that $\mu^{*}$ does not align with the oracle assignment rule, but does dictate that all individuals are assigned treatment (as they are under the oracle assignment rule);
$\kappa^{*}$ is also relatively large, as compared to its value in the previous experiments.

So as to make the assignment problem harder -- at least, what we understand to be harder -- we propose Experiments 9 and 10. 
These experiments follow Experiments 7 and 8 in how individual characteristics are generated, but differ slightly in how they generate the outcome.
Whilst Experiments 9 and 10 broadly follow Experiments 7 and 8 with respect to how the outcome is generated, they shift the process for the potential outcomes and, thereby, the oracle assignment rule.
Specifically, in Experiments 9 and 10, we assume that 
\begin{equation}
\label{EQ:BASELINE-910}
\arraycolsep=0pt
\begin{array}{lclclcl}
{\eta}&{}={}&(+1,286,+86,446,+14,008),&{}\hspace{15pt}{}&\sigma_{1}&{}={}&15,914,\\
{\alpha}&{}={}&(-668,+82,458,+18,804),&{}\hspace{15pt}{}&\sigma_{0}&{}={}&15,914,
\end{array}
\hspace{15pt}m=5\sigma_{1}=5\sigma_{0},
\end{equation}
such that the oracle policy is 
\begin{equation}
G
=
\{{x}\,:\,{x}\cdot{\beta}\geq0,{\beta}=(+0.098,+0.636,-0.765)\}.
\label{EQ:EXP-910-OPTIMAL}
\end{equation}
The effect of this change is to maintain the slope of the oracle assignment rule (i.e., the contribution of pre-programme earnings relative to years of education) but to alter the intercept. 
In other words, to narrow the difference in baseline average outcomes between the experimental and control groups.
In this regard, Experiments 9 and 10 are similar in intent to Experiment 4.

\begin{figure}[htb]
    \centering
    \caption{(Experiments 9--10) Variation in treatment propensity across individuals}
    \label{AFIG:PS-910}
    \includegraphics[width=\textwidth,height=3.5in]{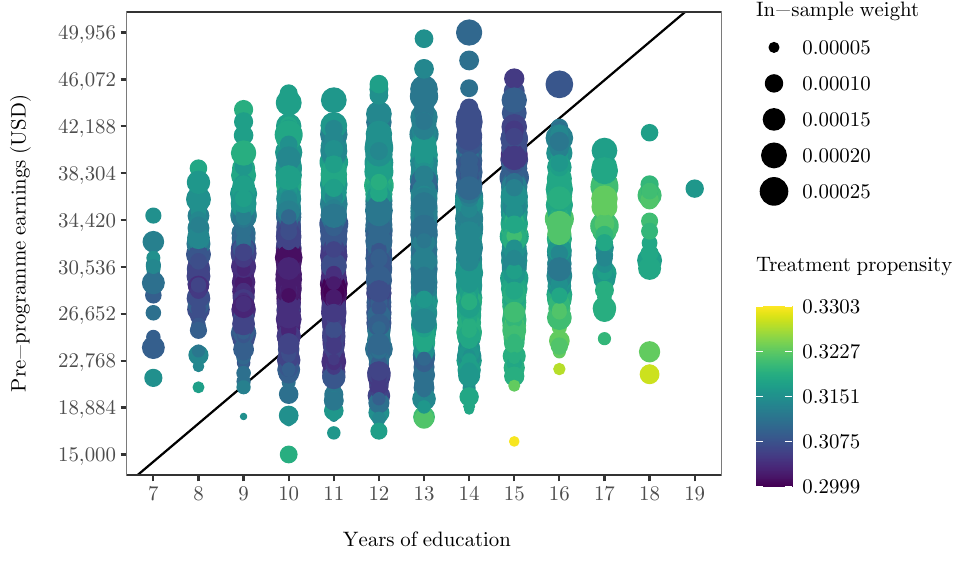}
        \includegraphics[width=\textwidth,height=3.5in]{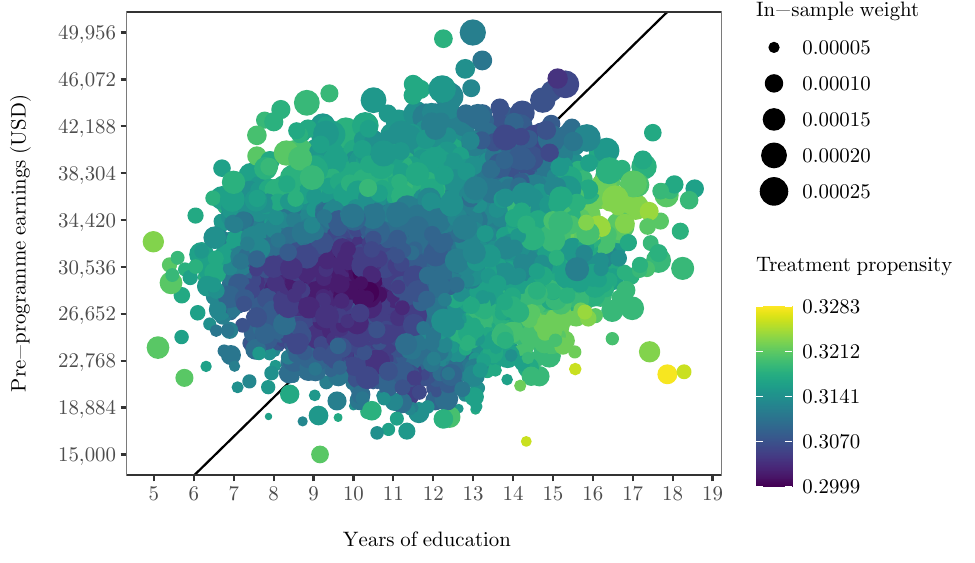}
    \caption*{\footnotesize\textit{Notes:}
    Simulated data generated using \cref{EQ:BASELINE-910} and a bivariate normal distribution.
    This figure illustrates the treatment propensity of individuals in simulated samples under the posterior assignment rule that is induced by $\left(\kappa^{*},\mu^{*}\right)$.
    Each point represents the individual characteristics of an individual or several individuals.
    For comparison, individuals to the left of the solid diagonal line are assigned treatment under the oracle assignment rule, which partitions the sample into two approximately equal-sized groups.} 
\end{figure}

\begin{figure}[htb]
    \centering
    \caption{(Experiments 9--10) Behaviour of the objective function at $\mu^{*}$ given variation in $\kappa$}
    \label{AFIG:KAPPA-910}
    \includegraphics[width=3in,height=3.5in]{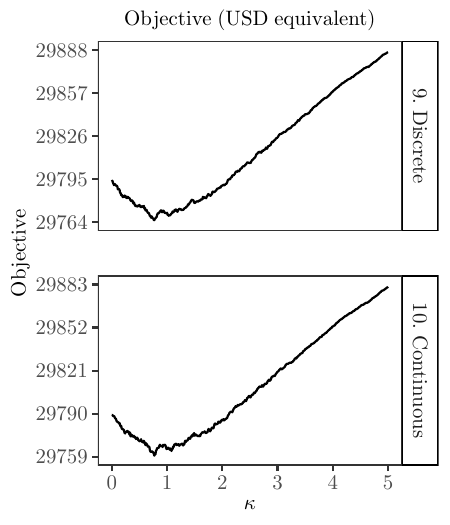}
        \includegraphics[width=3in,height=3.5in]{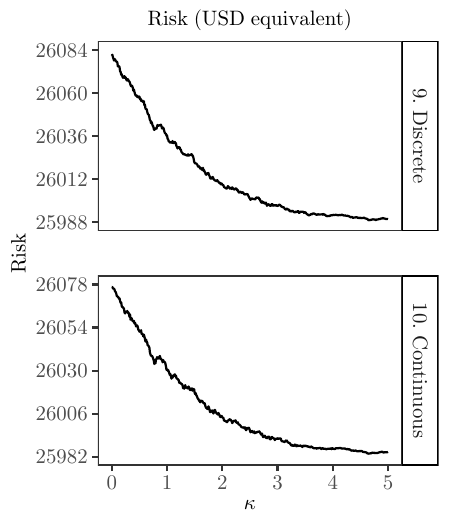}
    \caption*{\footnotesize\textit{Notes:}
    Simulated data generated using \cref{EQ:BASELINE-910} and a bivariate normal distribution.
    This figure illustrates how the shape of the objective function and its risk component at $\mu^{*}$ changes as $\kappa$ increases. }
\end{figure}

We plot the individual characteristics that we use in Experiments 9 and 10 in \cref{AFIG:PS-910} (\cpageref{AFIG:PS-910}).
The oracle assignment rule is designed to partition the sample into approximately two groups of equal size.
We plot the criterion (left-hand panel) and empirical welfare risk (right-hand panel) for Experiments 9 and 10 in \cref{AFIG:KAPPA-910} (\cpageref{AFIG:KAPPA-910}).
We draw attention to several differences between the results that we obtain for Experiments 9 and 10 versus Experiments 7 and 8.
First, we highlight the fall in the average propensity of treatment, as is apparent in \cref{AFIG:PS-910}.
Whereas in Experiments 7 and 8 (and the preceding experiments too), the propensity of treatment is close to one (i.e., everyone is assigned to the experimental group), here the propensity is closer to one third.
There is more randomisation.
Second, individuals that have a relatively high propensity of treatment as compared to their peers in Experiments 7 and 8 have a relatively low propensity of treatment in Experiments 9 and 10 (visually, the colours are inverted), as is apparent in \cref{AFIG:PS-910}.
That is, individuals whose characteristics are close to the mean values have a low propensity of treatment.
Third, the value of the concentration parameter decreases from above three to below one (i.e., the posterior distribution is less concentrated and more uniform), as is apparent in \cref{AFIG:KAPPA-910}.

We present the corresponding estimates of the parameters of the posterior distribution immediately below.

\begin{center}
\begin{tabular}{lcc}
         \hline
         \text{Experiment}& $\mu$ &$\kappa$\\
         
         \hline
         
         \text{9. Discrete}			& (-0.766,-0.470,-0.438) & 0.770 \\

    \text{10. Continuous}  				& (-0.766,-0.470,-0.438) & 0.770 \\
    
    \hline
\end{tabular}
\end{center}

Interestingly, the mean direction is, in both cases, almost the exact opposite of the corresponding mean direction in Experiments 7 and 8 (i.e., the mean direction is approximately the negative of the mean direction in Experiments 9 and 10.

\end{document}